\documentclass[reprint,twocolumn,showpacs,showkeys,preprintnumbers,amsmath,amssymb,aps,pre,floatfix]{revtex4-1}
\usepackage{graphicx}
\usepackage{dcolumn}
\usepackage{amssymb}
\usepackage{bm}
\usepackage[T1]{fontenc} 
\usepackage{color}
\newcommand{\LCO}{LCO}
\newcommand{\HCO}{HCO}
\newcommand{\PSCONO}{PS1}
\newcommand{\PSCOCO}{PS2}
\newcommand{\LCOA}{$\textrm{\LCO}_\textrm{A}$}%
\newcommand{\LCOB}{$\textrm{\LCO}_\textrm{B}$}%
\newcommand{\HCOA}{$\textrm{\HCO}_\textrm{A}$}%
\newcommand{\HCOB}{$\textrm{\HCO}_\textrm{B}$}%
\newcommand{\PSNDef}{$\textrm{\PSCONO}$}%
\newcommand{\PSNA}{$\textrm{\PSCONO}_\textrm{A}$}%
\newcommand{\PSNB}{$\textrm{\PSCONO}_\textrm{B}$}%
\newcommand{\PSNC}{$\textrm{\PSCONO}_\textrm{C}$}%
\newcommand{\PSCDef}{$\textrm{\PSCOCO}$}%
\newcommand{\PSCA}{$\textrm{\PSCOCO}_\textrm{A}$}%
\newcommand{\PSCB}{$\textrm{\PSCOCO}_\textrm{B}$}%
\newcommand{\PSCC}{$\textrm{\PSCOCO}_\textrm{C}$}%
\newcommand{\PSC}{$\textrm{\PSCOCO}^{*}$}%
\newcommand{\figwidth}{0.235\textwidth} 
\newcommand{\figwidthone}{0.37\textwidth} 
\begin{document}
\preprint{Submitted to: JOURNAL OF PHYSICS: CONDENSED MATTER}
\title{The effects of the next-nearest-neighbour density-density interaction\\
in the atomic limit of the extended Hubbard model}
%
\author{Konrad Kapcia}%
    \email[corresponding author; e-mail: ]{kakonrad@amu.edu.pl}%
\author{Stanis\l{}aw Robaszkiewicz}%
\affiliation{Electron States of Solids Division, Faculty of Physics, Adam Mickiewicz University, Umultowska 85, PL-61-614 Pozna\'n, Poland}
\date{January 22, 2011}
\begin{abstract}
We have studied the extended Hubbard model in the atomic limit. The Hamiltonian analyzed consists of the effective on-site interaction $U$ and the intersite density-density interactions $W_{ij}$ (both: nearest-neighbour and next-nearest-neighbour). The model can be considered
as a simple effective model of charge ordered insulators.
The phase diagrams and thermodynamic properties of this system have been determined within the variational approach, which treats the on-site interaction term exactly and the intersite interactions within the mean-field approximation.
Our investigation of the general case taking into account for the first time the effects of longer-ranged density-density interaction (repulsive and attractive) as well as possible phase separations shows that, depending on the values of the interaction parameters and the electron concentration, the system can exhibit not only several homogeneous charge ordered (CO) phases, but also various phase separated states (\mbox{CO-CO} and \mbox{CO-nonordered}).
One finds that the model considered exhibits very interesting multicritical behaviours and features, including among others bicritical, tricritical, critical-end and isolated critical points.
\end{abstract}
\maketitle
\section{Introduction}\label{sec:intro}
Electron charge orderings phenomena in strongly correlated electron systems are currently under intense investigations. Charge orderings (COs) are relevant to a~broad range of important materials, including manganites, cuprates, magnetite, several nickel, vanadium and cobalt oxides, heavy fermion systems (e.~g. \mbox{Yb$_4$As$_3$})  and numerous organic compounds \cite{IFT1998,GSS1985,F2006,VHN2001,SHF2004,IKM2004,MLC1990,GL2003,FT2006,DHM2001,DKC2001,RAA2005,QBK2004,RMR1987,MR1997,MRR1990,V1989,GK1994}.

Various types of COs have been also observed in a~great number of experimental systems with local electron pairing (for review see \cite{RMR1987,MR1997,MRR1990} and references therein), in particular in the compounds that contain cations in two valence states differing by $2e$ (on-site pairing) -- valence skipping, ``negative-$U$'' centers \cite{V1989}, and in the transition metal oxides showing intersite bipolarons e.~g. \mbox{Ti$_{4-x}$V$_x$O$_7$}, \mbox{WO$_{3-x}$}, with double charge fluctuations on the molecular (rather than atomic) units \mbox{[(Ti$^{4+}$-Ti$^{4+}$),(Ti$^{3+}$-Ti$^{3+}$)]}, etc.

COs are often found in broad ranges of electron doping (e.~g. doped manganites~\cite{GL2003,FT2006,DHM2001,DKC2001,RAA2005,QBK2004}, nickelates~\cite{IFT1998,GSS1985}, \mbox{Ba$_{1-x}$K$_{x}$BiO$_3$}~\cite{MRR1990,V1989,GK1994}). In several of these systems many experiments showed phase separations involving charge orderings~\cite{DHM2001,DKC2001,RAA2005,QBK2004}. The CO transitions at \mbox{$T>0$} take place either as first order or continuous.
Moreover, some of CO systems exhibit also a~tricritical behaviour (e.~g.~\mbox{(DI-DCNQI)$_2$Ag}~\cite{IKM2004}).


An important, conceptually simple model for studying correlations and for description of charge orderings (and various other types of electron orderings) in narrow energy band systems is
the extended Hubbard model taking into account both the on-site ($U$) and the intersite ($W_{ij}$) density-density interactions  (the \mbox{$t$-$U$-$W_{ij}$} model~\mbox{\cite{MRR1990,MR1992,TSB2004,M2005,MYI2006,SMY2006,R1994}}).

In this paper we focus on the atomic limit ($t_{ij}=0$ limit) of the \mbox{$t$-$U$-$W_{ij}$} model.
The Hamiltonian considered has the following form:
\begin{eqnarray}
\label{row:1}
\hat{H} &= & U\sum_i{\hat{n}_{i\uparrow}\hat{n}_{i\downarrow}} + \frac{W_{1}}{2}\sum_{\langle i,j\rangle_1}{\hat{n}_{i}\hat{n}_{j}}+ \nonumber\\
&+& \frac{W_{2}}{2}\sum_{\langle i,j\rangle_2}{\hat{n}_{i}\hat{n}_{j}} - \mu\sum_{i}{\hat{n}_{i}},
\end{eqnarray}
where $\hat{c}^{+}_{i\sigma}$ denotes the creation operator of an electron with spin \mbox{$\sigma=\uparrow,\downarrow$} at the site $i$, \mbox{$\hat{n}_{i}=\sum_{\sigma}{\hat{n}_{i\sigma}}$}, \mbox{$\hat{n}_{i\sigma}=\hat{c}^{+}_{i\sigma}\hat{c}_{i\sigma}$}. $\mu$~is the chemical potential.
\mbox{$\sum_{\langle i,j\rangle_{m}}$} indicates the sum over nearest-neighbour (\mbox{$m=1$}) and next-nearest-neighbour (\mbox{$m=2$}) sites $i$ and $j$ independently. $z_m$ denotes the number of $m$-th neighbours.
$U$ is the on-site density interaction,
$W_{1}$ and $W_{2}$ are the intersite density-density interactions between nearest neighbours (nn)
and next-nearest neighbours (nnn), respectively.
The chemical potential $\mu$ depends on the concentration
of electrons:
\begin{equation}\label{row:2}
n = \frac{1}{N}\sum_{i}{\left\langle \hat{n}_{i} \right\rangle},
\end{equation}
with \mbox{$0\leq n \leq 2$} and $N$ -- the total number of lattice sites. $\langle \hat{n}_i \rangle$ denotes the average value of the $\hat{n}_i$ operator.

The model~(\ref{row:1}) can be considered as a simple model of charge ordered insulators.
The interactions $U$ and $W_{ij}$ can be treated as the effective ones and assumed to include all the possible contributions and renormalizations like those coming from the strong electron-phonon coupling or from the coupling between electrons and other electronic subsystems in solid or chemical complexes. In such a general case arbitrary values and signs of $U$ and $W_{ij}$ are important to consider.

Notice that  the model (\ref{row:1}) can be viewed as the classical gas with four possible states at each site and it is equivalent to a special kind of the Blume-Capel model  i.~e.~the \mbox{$S=1$} Ising model with single-ion  anisotropy and eigenvalue `zero' doubly degenerate in an effective magnetic field given self-consistently by a~value of fixed magnetization \cite{MRC1984,BJK1996,P2006}.

In the analysis we have adopted a variational approach (VA) which treats the on-site interaction $U$ exactly and the intersite interactions $W_{ij}$ within the mean-field approximation (MFA).

Within the VA the phase diagrams of (\ref{row:1}) have been investigated till now for the  case \mbox{$W_2=0$}~\mbox{\cite{B1971,R1979,MRC1984}} and the stability conditions of states with phase separation have not been discussed. Some preliminary results for the case \mbox{$W_2\neq0$} have been presented by us in~\cite{KKR2010}.

We perform extensive study of the phase diagrams and thermodynamic properties of the model~(\ref{row:1}) within VA for arbitrary electron concentration $n$, arbitrary strength of the on-site interaction $U$  and the nn repulsion \mbox{$W_1>0$}, taking into account the effects of  interactions between nnn $W_2$ (repulsive and attractive).
Our comprehensive investigation of the general case finds that, depending on the values of the interaction parameters and the electron concentration, the system can exhibit the charge ordered and nonordered homogeneous phases as well as (for attractive $W_2$) at least two types of phase separation involving charge orderings. Transitions between different states and phases can be continuous and discontinuous, what implies existence of different critical points on the phase diagrams. We present detailed results concerning the evolution of phase diagrams as a~function of the interaction parameters and the electron concentration.

Our studies of the Hamiltonian~(\ref{row:1}) are exact for attractive $W_2$ in the limit of infinite dimensions. They are important as a~test and a~starting point
for a perturbation expansion in powers of the hopping $t_{ij}$ and as a benchmark for various approximate approaches (like dynamical mean field approximation, which is exact theory for fermion system in the limit of infinite dimensions for \mbox{$t_{ij}\neq0$} \cite{TSB2004}) analyzing the corresponding finite bandwidth models. They can also be useful in a~qualitative analysis of experimental  data for real narrow bandwidth materials in which  charge orderings phenomena are observed.

In the limit \mbox{$W_2=0$} the model~(\ref{row:1}) has been analyzed in detail (for review see~\cite{MRC1984,MM2008} and references therein). In particular, the exact solutions were obtained for the one-dimensional (\mbox{$d=1$}) case (\mbox{$T\geq0$}) employing the method based on the equations of motion and Green's function formalism~\cite{MM2008} or the transfer-matrix method~\cite{BP1974,TK1974}.
In~\cite{BJK1996} the phase diagram of~(\ref{row:1}) as a~function of $\mu$ for \mbox{$W_2=0$} has been derived at  \mbox{$T=0$} and confirmed in~\cite{FRU2001}. These studies were based on the Pirogov and Sinai methods~\cite{PS1975}.
In two dimensional case \mbox{$d=2$} (\mbox{$W_2\neq 0$}) exact ground state diagrams as a function of $\mu$ have also been obtained~\cite{J1994} using the metod of constructing ground state phase diagrams by the reflection positivity property with respect to reflection in lattice planes.

A number of numerical simulation has also been done for \mbox{$W_2=0$}.  In particular, the critical behaviour near the tricritical point have been analyzed using Monte Carlo (MC) simulation and MFA~\cite{MYI2006}. A~study of the model in finite temperatures using MC simulation has also been done for a square lattice (\mbox{$d=2$})~\cite{P2006,GP2008}. In particular, the possibility of phase separation and formation of stripes in finite systems was evidenced there.

In the following we will restrict ourselves to the case of repulsive \mbox{$W_1>0$}, which favours charge orderings, and \mbox{$z_1W_1>z_2W_2$}. For the sake of simplicity we consider mainly two-sublattice orderings
on the alternate lattices, i.~e. the lattices consisting of two interpenetrating sublattices (every nearest neighbour of every site in one sublattice is a site in the other sublattice), such as for example simple cubic (SC) or body-center cubic lattices. Some preliminary studies of ground state beyond the two-sublattice assumption at half-filling (\mbox{$n=1$}) are also performed, which show that for repulsive $W_2$ there is possibility of occurrence of the multi-sublattice orderings.

The paper is organized as follows. In section~\ref{sec:ham} we describe the metod used in this work. There are also derived explicit formulas for the free energies of homogeneous phases and states with phase separation as well as equations determining the charge-order parameters and the chemical potential in homogeneous phases. In section~\ref{sec:GS} we analyze the properties of the system at zero temperature and present ground state diagrams. Section~\ref{sec:FTW2} is devoted to the study of the finite temperature phase diagrams for \mbox{$W_2\geq0$} and \mbox{$W_2<0$}. Some particular temperature dependencies of the charge-order parameter are discussed in section~\ref{sec:termodynamics}. Section~\ref{sec:beyond} contains ground state  results for half-filling beyond two-sublattice assumption.  Finally, section~\ref{sec:conclusions} reports the most important conclusions and supplementary discussion including the validity of the approximation used and the comparison with real materials. The appendix presents  explicit expressions of site-dependent self-consistent VA equations.

\section{The method}\label{sec:ham}

The free energy of the system and the self-consistent equations for the average number of electrons on sites are derived within site-dependent VA in the Appendix.
Restricting analysis to the two-sublattice orderings the explicit formula for the free energy per site
obtained in the VA has the following form
\begin{equation}\label{row:freeenergy}
f(n)=\frac{F}{N}=\mu n -\frac{1}{2}W_0n^2-\frac{1}{2}W_Qn_Q^2-\frac{1}{2\beta}\ln\left[Z_A Z_B\right],
\end{equation}
where
\begin{eqnarray*}
 Z_{\alpha} & = & 1+2\exp[\beta(\mu-\psi_{\alpha})] + \exp[\beta( \mu - 2 \psi_{\alpha}-U)],\\
 \psi_A  & = & nW_0 + n_Q W_Q, \qquad  \psi_B  =  nW_0 - n_Q W_Q, \\
 W_0 & = & z_1W_1+z_2W_2,  \qquad   W_Q = -z_1W_1+z_2W_2,
\end{eqnarray*}
and $\beta=1/(k_BT)$.
The charge-order parameter is defined as \mbox{$n_Q=(1/2)(n_A-n_B)$}, where \mbox{$n_{\alpha}=\frac{2}{N}\sum_{i\in \alpha}{\langle \hat{n}_i \rangle}$} is the average electron density in a sublattice \mbox{$\alpha=A,B$}.

The condition for the electron concentration (\ref{row:2}) and a~minimization of $f(n,T)$ with
respect to the charge-order parameter lead to a set of self-consistent equations (for homogeneous phases):
\begin{eqnarray}
\label{row:set1}n = (1/2)\left(n_A+n_B\right),\\
\label{row:set2}n_Q = (1/2)\left(n_A-n_B\right),
\end{eqnarray}
where
\begin{equation*}
n_\alpha = \frac{2}{Z_{\alpha}}\left\{ \exp{\left[\beta(\mu - \psi_\alpha )\right]}+ \exp{\left[\beta(2\mu - 2 \psi_\alpha -U)\right]}\right\}.
\end{equation*}
The double occupancy per site defined as
\mbox{$D = \frac{1}{N}\sum_{i}\left\langle \hat{n}_{i\uparrow}\hat{n}_{i\downarrow}\right\rangle$} has the following form:
\begin{equation}
D = (1/2)\left( D_A + D_B \right),
\end{equation}
where \mbox{$D_\alpha=\exp{[\beta(2\mu - 2 \psi_\alpha -U)]}/Z_\alpha$}.

The equations (\ref{row:set1})--(\ref{row:set2})  are solved numerically for \mbox{$T\geq0$} and we obtain $n_Q$
and $\mu$ when $n$ is fixed.
The charge-ordered (CO) phase  is characterized by non-zero value of $n_Q$, whereas \mbox{$n_Q=0$} in the non-ordered (NO) phase.

Let us notice that the free energy (\ref{row:freeenergy}) is an even function of $n_Q$ so we can restrict ourselves to solutions of the set (\ref{row:set1})--(\ref{row:set2}) in the range \mbox{$0\leq n_Q \leq 1$}. It is the result of the equivalence of two sublattices. Moreover, (\ref{row:set2}) is only the necessary condition for an~extremum of (\ref{row:freeenergy}) thus the solutions of (\ref{row:set1})--(\ref{row:set2}) can correspond to a~minimum or a~maximum (or a~point of inflection) of (\ref{row:freeenergy}). In addition the number of minimums can be larger than one, so it is very important to find the solution which corresponds to the global minimum of (\ref{row:freeenergy}).

Phase separation (PS) is a state in which two domains with different electron concentration: $n_+$ and $n_-$ exist in the system
(coexistence of two homogeneous phases). The free energies of the PS states are calculated from the expression:
\begin{equation}\label{row:freeenergyPS}
f_{PS}(n_{+},n_{-}) = m f_{+}(n_{+}) + (1-m) f_{-}(n_{-}),
\end{equation}
where $f_{\pm}(n_{\pm})$ are values of a free energy at $n_{\pm}$ corresponding to the
lowest energy homogeneous solution and
\mbox{$m  =(n - n_-)/(n_+ - n_-)$}
is a fraction of the system with a charge density $n_+$ (\mbox{$n_-<n<n_+$}). The minimization of (\ref{row:freeenergyPS}) with respect to $n_+$ and $n_-$ yields the equality between the chemical potentials in both domains:
\begin{equation}\label{row:PS1}
\mu_+(n_+)=\mu_-(n_-)
\end{equation}
(chemical equilibrium)
and the following equation (so-called Maxwell's construction):
\begin{equation}\label{row:PS2}
\mu_+(n_+)=\frac{f_{+}(n_{+})-f_{-}(n_{-})}{n_{+}-n_{-}}.
\end{equation}
In the PS states the chemical potential \mbox{$\mu=\mu_+(n_+)=\mu_-(n_-)$} is independent of the electron concentration, i.e. \mbox{$\partial \mu/\partial n=0$}.

In the model considered only the following types of PS states can occur: (i)~{\PSNDef} is a coexistence of CO and NO phases and (ii)~{\PSCDef} is a coexistence of two CO phases with different concentrations.

In the paper we have used the following convention. A~second (first) order transition is a~transition between homogeneous phases  with a~(dis-)continuous change of the order parameter at the transition temperature. A~transition between homogeneous phase and PS state is symbolically named as a~``third order'' transition. During this transition a~size of one domain in the PS state decreases continuously to zero at the~transition temperature.  We have also distinguished a~second (first) order transition between two PS states, at which a~\mbox{(dis-)continuous} change of the order parameter in one of domains takes place. In the both cases the order parameter in the other domain changes continuously.

Second order transitions are denoted by solid lines on phase diagrams, whereas dotted and dashed curves denote first order and ``third order'' transitions, respectively.

The phase diagrams obtained are symmetric with respect to half-filling because of the particle-hole symmetry of the hamiltonian (\ref{row:1}) \cite{MRR1990,S1972,MM2008}, so the diagrams will be presented only in the range \mbox{$0\leq n\leq 1$}.

\section{The ground state}\label{sec:GS}

\subsection{\mbox{$W_2>0$}}

In the case of nnn repulsion, i.~e. \mbox{$0<k<1$} (\mbox{$k=z_2W_2/z_1W_1$}),  the system at \mbox{$T=0$} can exhibit two types of CO: ``\emph{high} CO'' (\HCO), involving the on-site pairing of electrons and ``\emph{low} CO'' (\LCO), which is the ordering without on-site pairs.

\begin{figure}
        \centering
        \includegraphics[width=\figwidth]{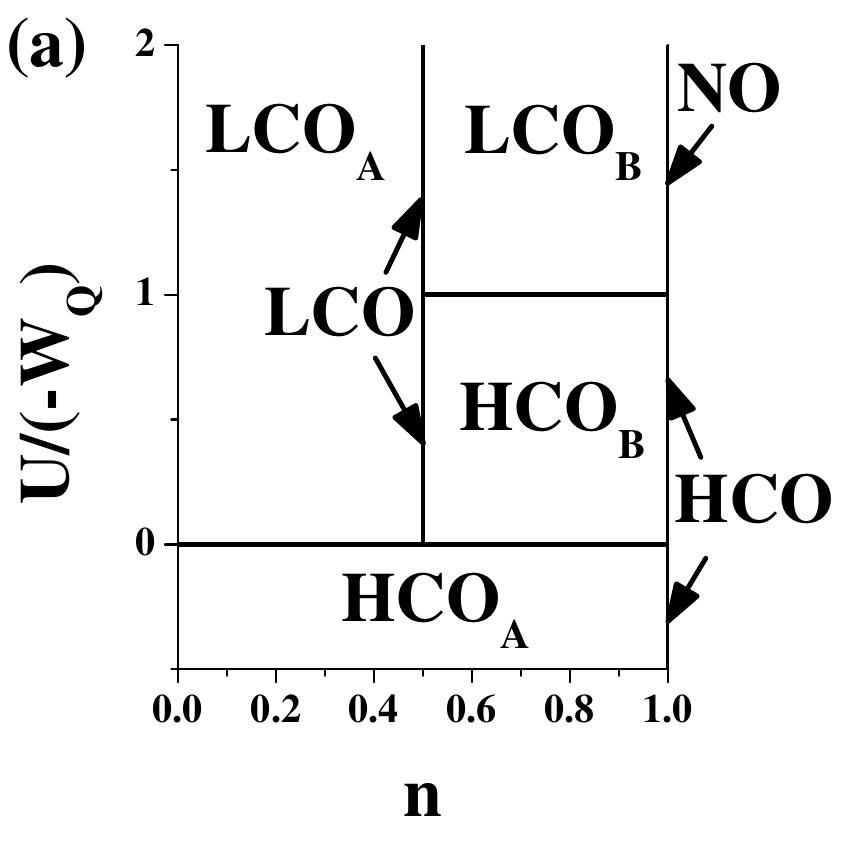}
        \includegraphics[width=\figwidth]{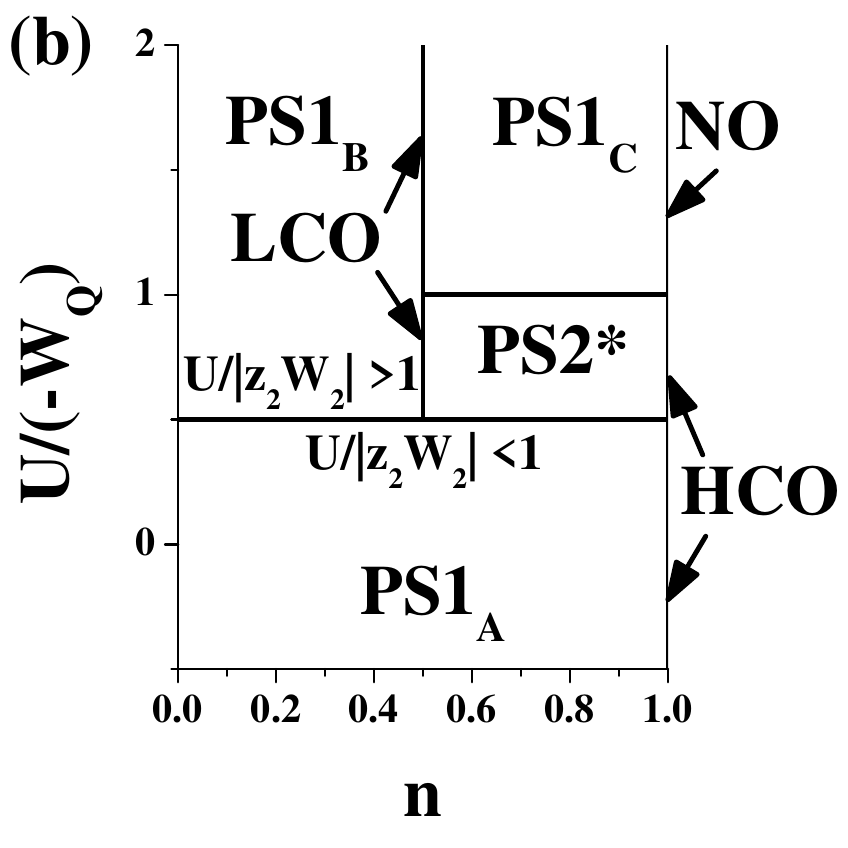}
        \caption{Ground state phase diagrams $U/(-W_Q)$~vs.~$n$: (a)~for \mbox{$1>k>0$} and (b)~for \mbox{$k<0$} (\mbox{$k=-1$}). Details in text and Table~\ref{tab:table2}.}
        \label{rys:GS}
\end{figure}

For this case the ground state (GS) diagram derived within VA as a function of $n$ and $U/(-W_Q)$  is shown in figure~\ref{rys:GS}a.
At \mbox{$T=0$} {\HCO} can be stable phase only if \mbox{$U/(-W_Q)<1$} and {\LCO} if \mbox{$U>0$}. For \mbox{$0<U/(-W_Q)<1$} both types of order can be realized depending on $n$.
In the GS, one obtains the following results for the ordering parameter $n_Q$, the chemical potential $\mu$ and the double occupancy per site $D$:
(i)~{\LCO} phases:
for {\LCOA} (only sublattice $A$ is filled by electrons without double occupancy \mbox{$n_A=2n$}, sublattice $B$ is empty): \mbox{$n_Q=n$}, \mbox{$\mu=2z_2W_2n$}, \mbox{$D=0$}
and for {\LCOB} (every site in sublattice $A$ is singly occupied \mbox{$n_A=1$}, whereas \mbox{$n_B=2n-1$}): \mbox{$n_Q=1-n$}, \mbox{$\mu=2z_2W_2n-W_Q$}, \mbox{$D=0$},
(ii)~{\HCO} phases (electrons are located only in sublattice $A$, \mbox{$n_A=2n$}, \mbox{$n_B=0$}):
for {\HCOA} (every site in sublattice $A$ is doubly occupied):  \mbox{$n_Q =n$}, \mbox{$\mu=U/2+2z_2W_2n$}, \mbox{$D=n/2$}
and for {\HCOB} (at every site in sublattice $A$ at least one electron is located): \mbox{$n_Q=n$}, \mbox{$\mu=U+2z_2W_2n$}, \mbox{$D=n-1/2$}.
For \mbox{$n=1$} the transition at \mbox{$U/(-W_Q)=1$} is from the {\HCO} (\mbox{$n_Q=1$}, \mbox{$D=0.5$}) to the NO (Mott state, \mbox{$n_Q=0$}, \mbox{$D=0$}). In both phases at \mbox{$n=1$} the chemical potential is \mbox{$\mu=U/2+W_0=U/2+2z_2W_2-W_Q$}. At quarter-filling (\mbox{$n=0.5$}) in {\LCO} (\mbox{$n_Q=0.5$}, \mbox{$D=0$}) the value of the chemical potential is \mbox{$\mu=U/2+z_2W_2$} for \mbox{$0<U/(-W_Q)<1$} and \mbox{$\mu=(1/2)W_0$} for \mbox{$U/(-W_Q)>1$}.

For fixed $U/(-W_Q)$ the chemical potential $\mu$ changes discontinuously (except \mbox{\LCO--\HCOB} and \mbox{\LCO--\LCOB}), whereas for fixed $n$ it is continuous at the phase boundaries. The double occupancy $D$ changes continuously at transitions with fixed $U/(-W_Q)$, while transitions with fixed $n$ are associated with discontinuous change of $D$ (except \mbox{\LCO--\LCO} for \mbox{$n=1/2$} and \mbox{$U/(-W_Q)=1$}). Apart from \mbox{\HCO--NO} (for \mbox{$n=1$}) and \mbox{\HCOB--\LCOB} transitions, the charge-order parameter $n_Q$ changes continuously in all transitions.

\begin{table}
    \caption{\label{tab:table2}PS states occurring in the ground state.}
    \begin{ruledtabular}
        \begin{tabular}{llclc}
            State & Domain & $n_+$ & Domain & $n_-$\\
            \hline\hline
            \PSNA & \HCO & $1$ & NO & $0$ \\
            \PSNB  & \LCO\footnotemark[1] & $0.5$ & NO & $0$\\
            \PSNC  & NO & $1$ & \LCO\footnotemark[1] & $0.5$\\
            \PSC  & \HCO & $1$ & \LCO\footnotemark[1] & $0.5$\\
        \end{tabular}
    \end{ruledtabular}
    \footnotetext[1]{The value of $\mu$ in the homogeneous phase is irrelevant here.}
\end{table}

\subsection{\mbox{$W_2<0$}}

In the GS diagram as a function of $n$ for \mbox{$W_2<0$} (figure~\ref{rys:GS}b) one finds also simple linear boundaries between various states but in this case the free energies of PS states are lower than those of homogeneous phases, apart from particular values of concentration: (i)~for \mbox{$n=0.5$} and \mbox{$U/|z_2W_2|>1$} the {\LCO} with \mbox{$n_Q=0.5$} is stable, (ii)~for \mbox{$n=1$} the {\HCO} with \mbox{$n_Q=1$} if \mbox{$U/(-W_Q)<1$} and the NO with \mbox{$n_Q=0$} if \mbox{$U/(-W_Q)>1$} are stable. One can also check that the first derivative of chemical potential $\partial\mu/\partial n$ in homogeneous phases is negative (apart from the ranges mentioned above) what implies that these phases are not stable. Definitions of all PS states occurring in the GS are collected in Table~\ref{tab:table2}. For \mbox{$U/|z_2W_2|<1$} only the {\PSNA} state (with \mbox{$D=n/2$}, \mbox{$\mu=U/2+z_2W_2$}) occurs. When \mbox{$0.5<n<1$}, \mbox{$U/|z_2W_2|>1$} and \mbox{$U/(-W_Q)<1$} the {\PSC} state with \mbox{$D=n-1/2$} and \mbox{$\mu=U+(3/2)z_2W_2$} is stable. For \mbox{$U/|z_2W_2|>1$} and \mbox{$n<0.5$} the {\PSNB} with \mbox{$D=0$} and \mbox{$\mu=z_2W_2/2$} has the lowest energy, whereas for \mbox{$U/|W_Q|>1$}  and \mbox{$0.5<n<1$} the {\PSNC} with \mbox{$D=0$} and \mbox{$\mu=(3/2)z_2W_2-W_Q$} is stable.

All transitions in GS for \mbox{$W_2<0$} are associated with discontinuous change of the chemical potential $\mu$. The double occupancy $D$ changes continuously at transitions with fixed $U/(-W_Q)$, while transitions with fixed $n$ are connected with discontinuous change of $D$ (except {\LCO}--{\LCO} for \mbox{$n=1/2$} and \mbox{$U/(-W_Q)=1$}).

Notice that for \mbox{$W_1>0$} and \mbox{$W_2<0$} the condition \mbox{$U/(-W_Q)=1$} implies that \mbox{$U/|z_2W_2|>1$}, so the line \mbox{$U/(-W_Q)=1$} is above the line \mbox{$U/|z_2W_2|=1$} and the GS phase diagram for any \mbox{$W_2<0$} has always the form shown in figure~\ref{rys:GS}b.

One should stress that for \mbox{$W_2<0$} the phase stability condition is not fulfilled (i.~e. \mbox{$\partial\mu/\partial n <0$}) in homogeneous phases except \mbox{$n=0.5$} for \mbox{$U/|z_2W_2|>1$} and \mbox{$n=1$}. It means that the homogeneous phases are not stable.

For \mbox{$W_2=0$} the free energies of homogeneous phases and PS states are degenerated at \mbox{$T=0$} and for such a~case \mbox{$\partial \mu / \partial n =0$} in homogeneous phases. This degeneration is removed in any finite temperatures and at \mbox{$T>0$} homogeneous phases have the lowest energy.
Our diagrams obtained for \mbox{$W_2=0$} are consistent with results presented in~\cite{MRC1984}.

\section{Finite temperatures}\label{sec:FTW2}

The behaviours of the system for repulsive \mbox{$W_2>0$} and attractive \mbox{$W_2<0$} are qualitatively different.

One obtains from numerical analysis that, for \mbox{$z_1W_1>z_2W_2\geq0$} taking into account only two-sublattice orderings, the PS states are unstable at any \mbox{$T>0$} and the homogeneous phases are stable (\mbox{$\partial\mu/\partial n>0$} for any \mbox{$T>0$}, even for \mbox{$W_2=0$}). The
finite temperature phase diagrams have the forms determined in~\cite{MRC1984},
with the replacements: \mbox{$z_1W_1 \rightarrow -W_Q >0$}. Thus we describe obtained  results in this case shortly, directing  the reader for detailed analyses to~\cite{MRC1984}. One can conclude that transition temperatures between homogeneous phases decrease with increasing \mbox{$W_2>0$} and $U/(-W_Q)$.

\begin{figure*}
        \centering
        \includegraphics[width=\figwidthone]{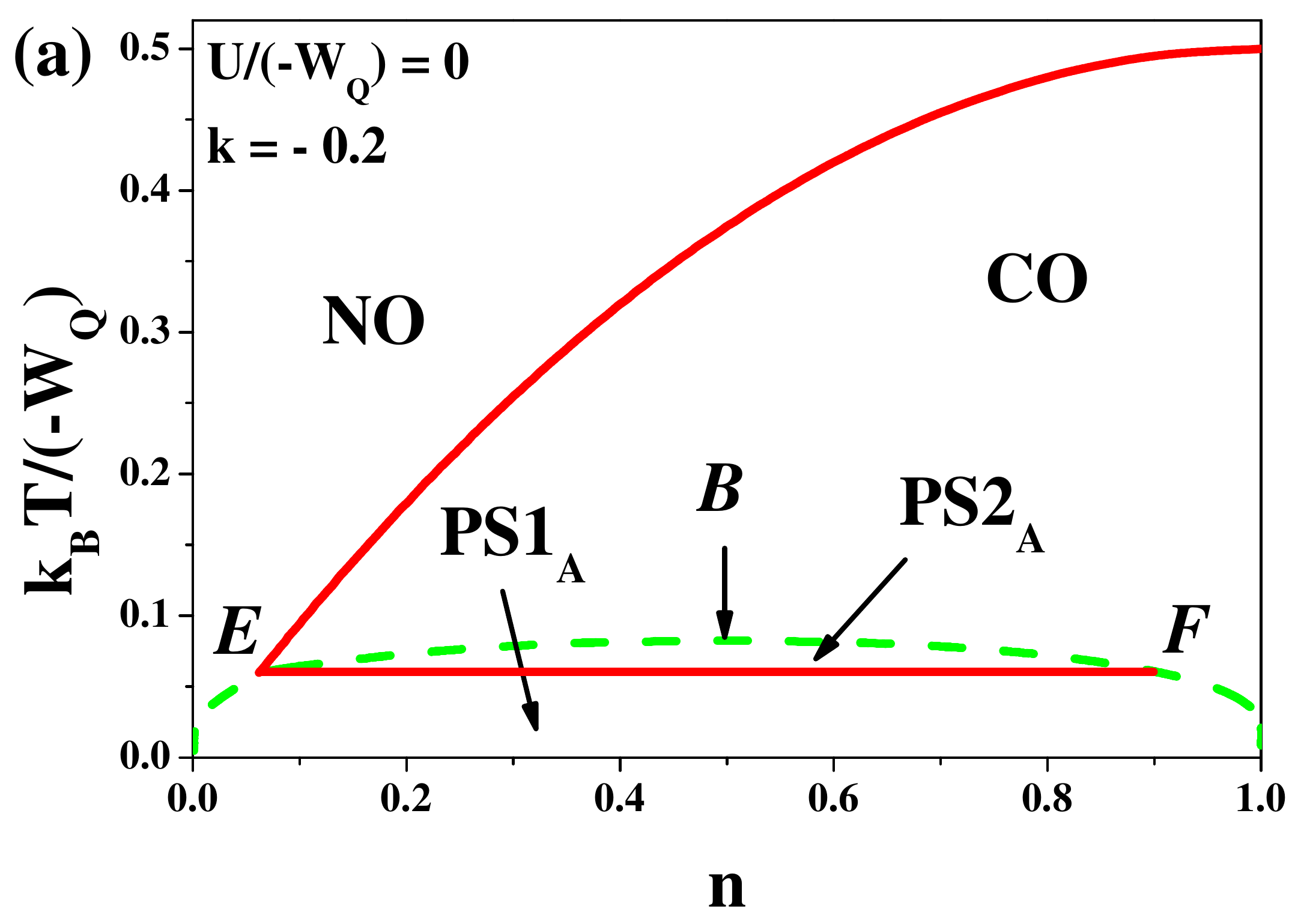}
        \includegraphics[width=\figwidthone]{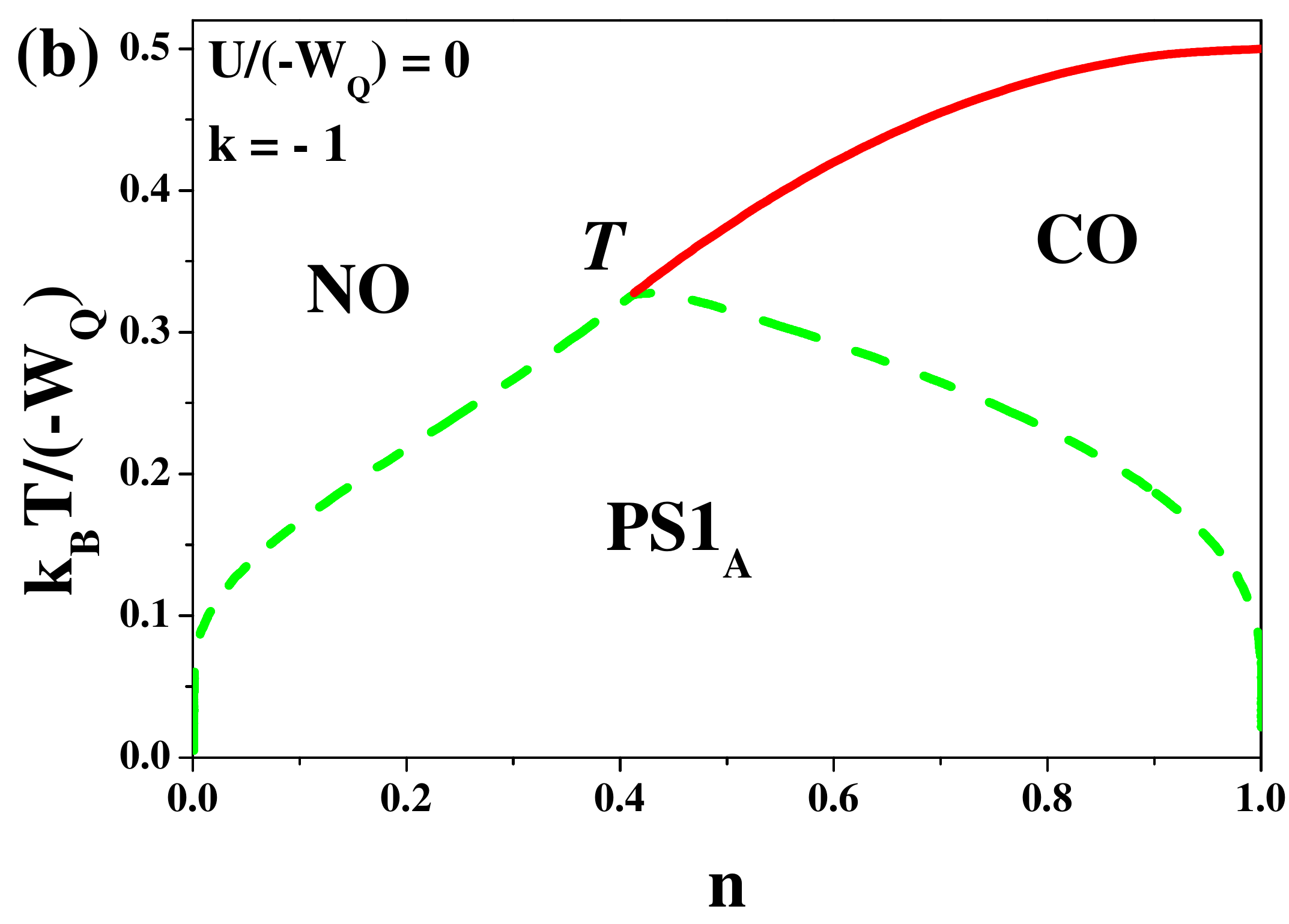}
        \caption{(Color online) Phase diagrams $k_BT/(-W_Q)$~vs.~$n$ for \mbox{$U/(-W_Q)=0$}, \mbox{$W_1>0$} and different values of \mbox{$k=z_2W_2/z_1W_1$}: \mbox{$k=-0.2$}~(a) and \mbox{$k=-1$}~(b). Solid and dashed lines indicate second order and ``third order'' boundaries, respectively.}
        \label{rys:U0}
\end{figure*}

\begin{figure*}
        \centering
        \includegraphics[width=\figwidthone]{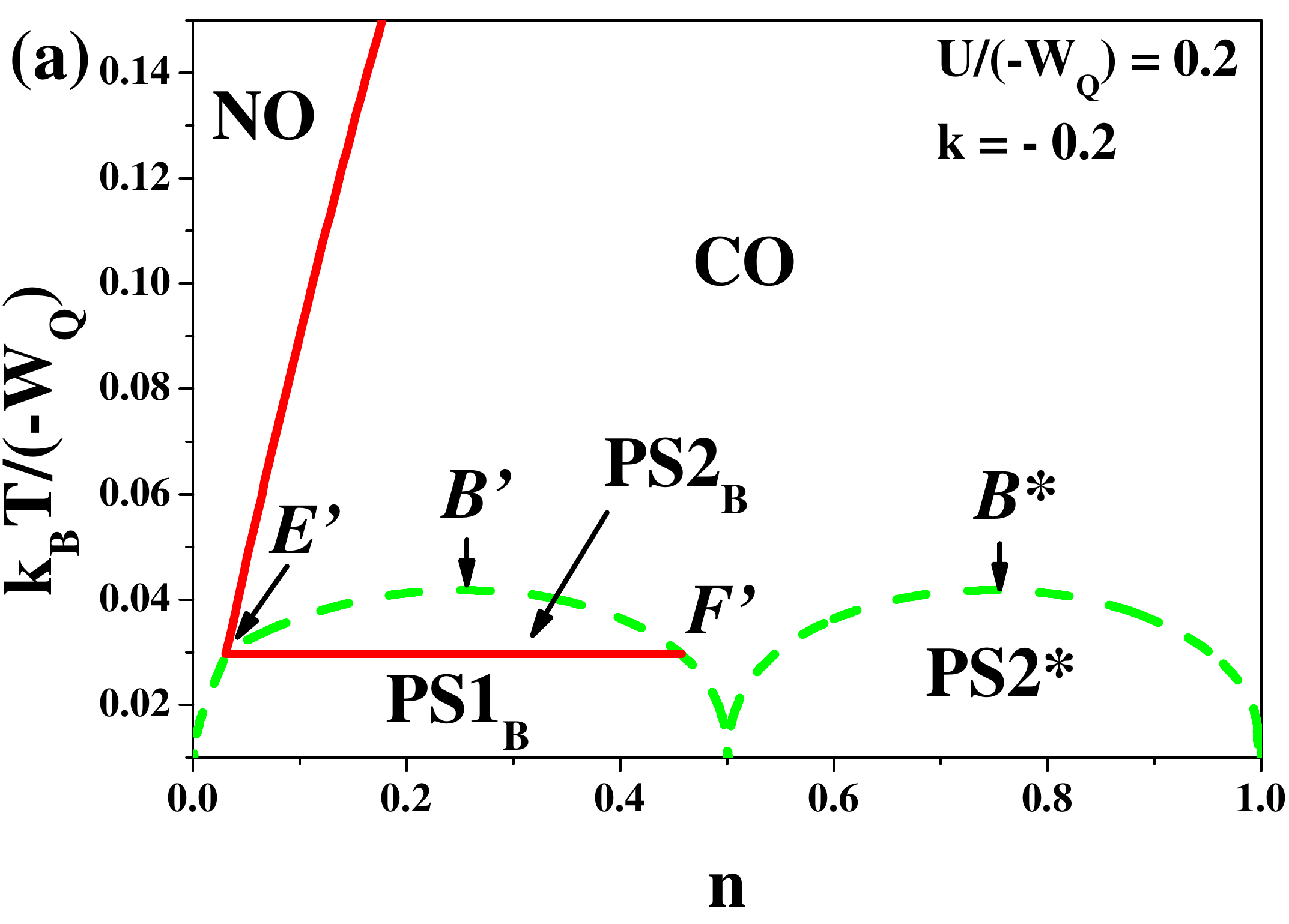}
        \includegraphics[width=\figwidthone]{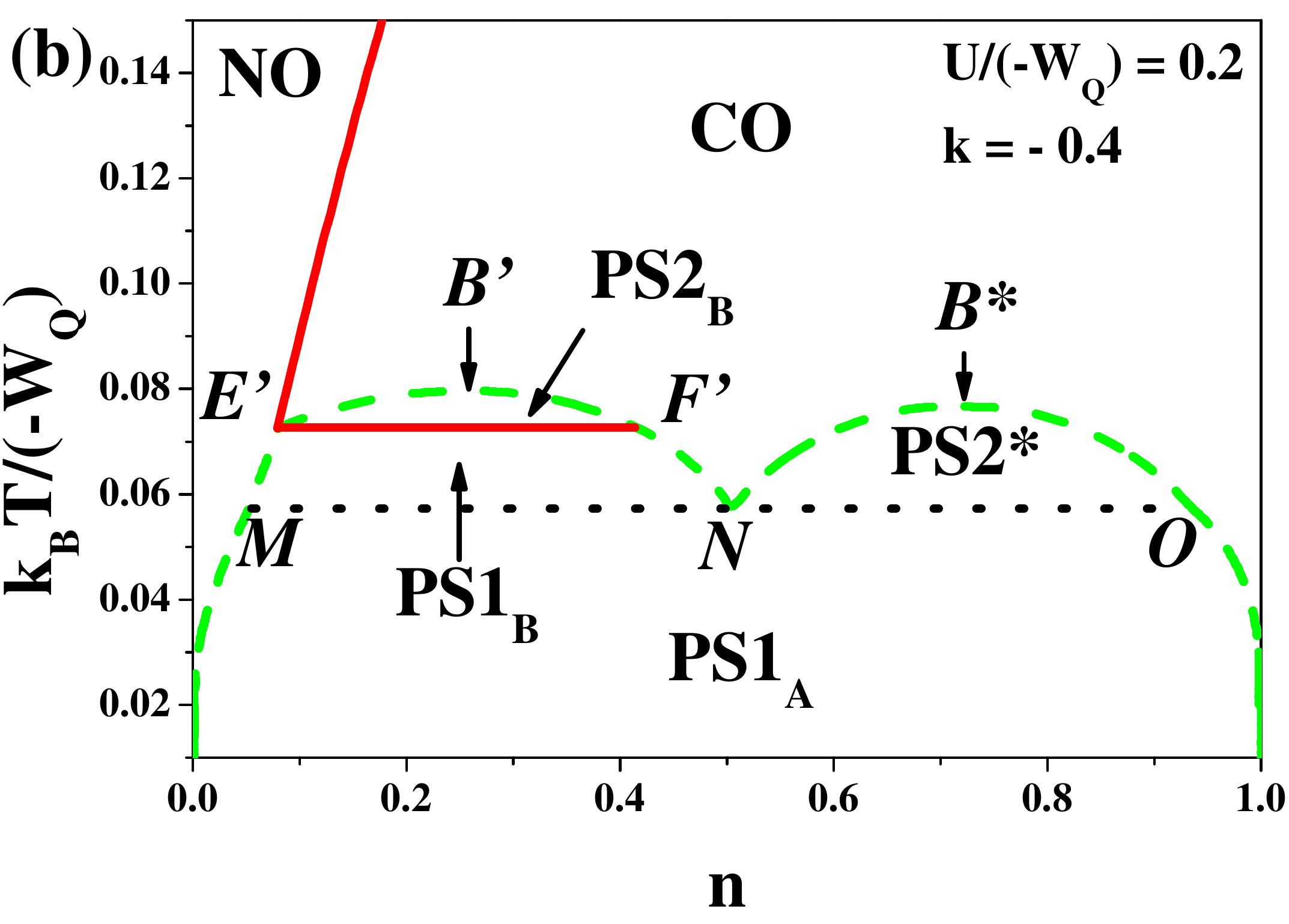}\\
        \includegraphics[width=\figwidthone]{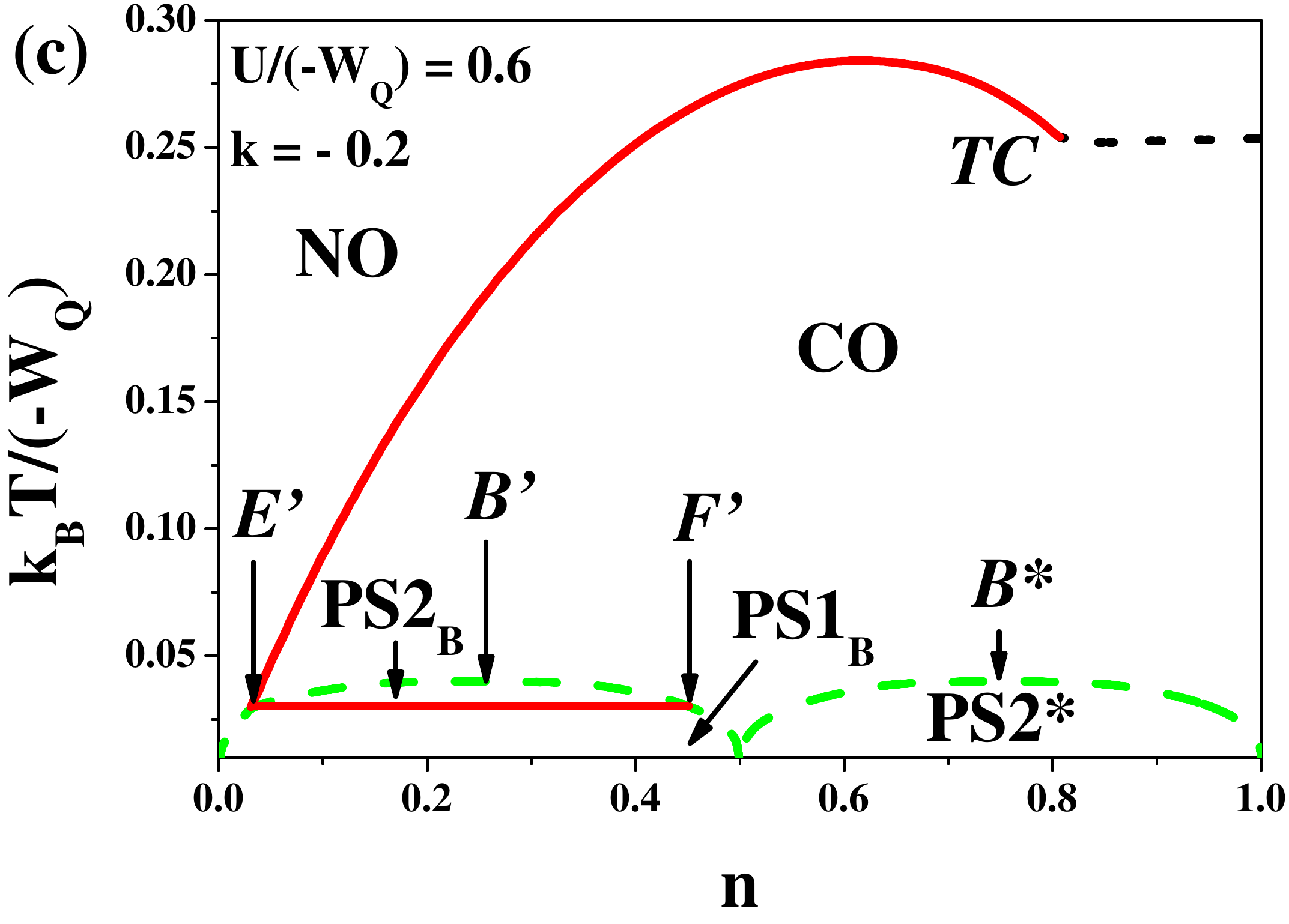}
        \includegraphics[width=\figwidthone]{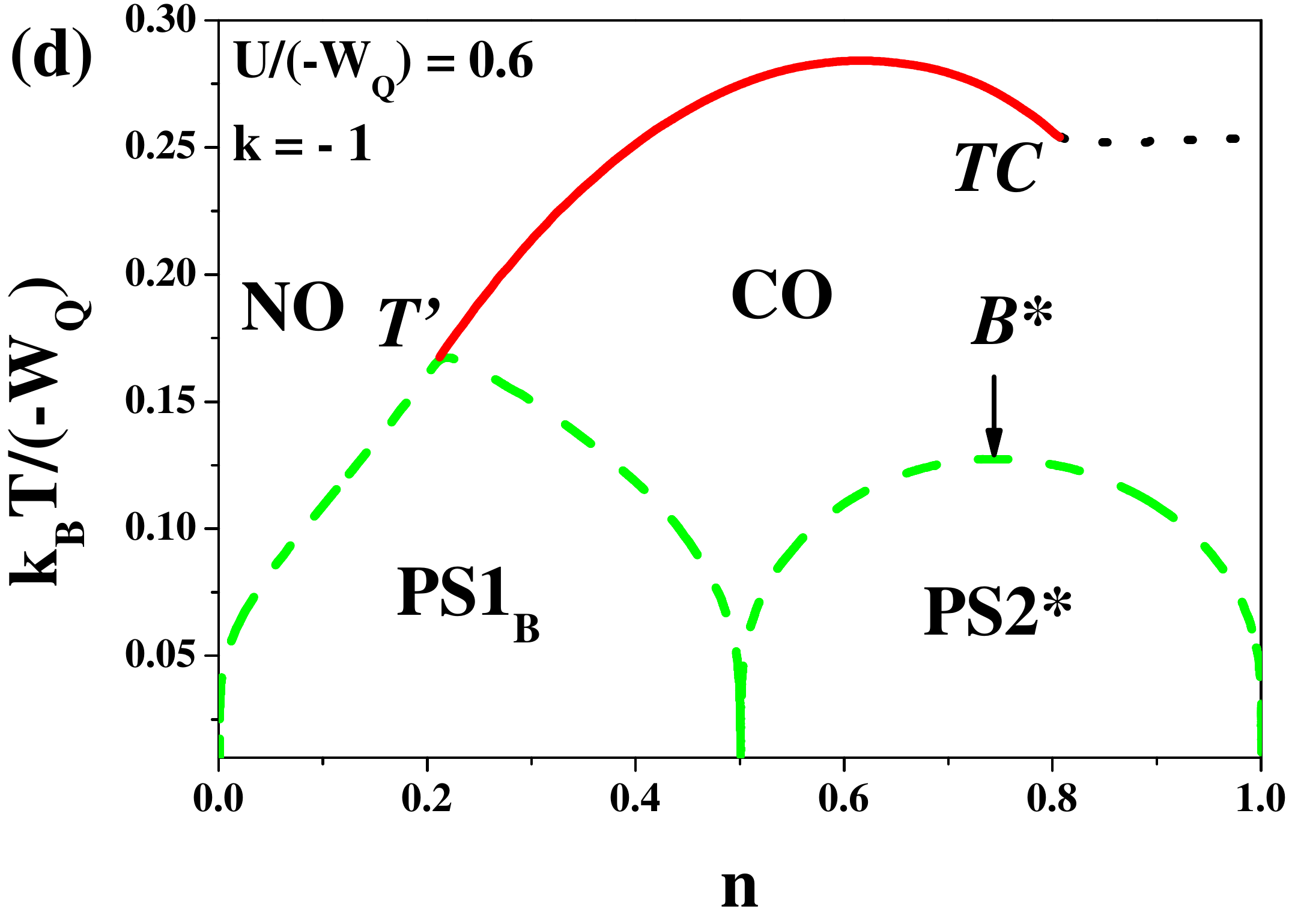}
        \caption{(Color online) Phase diagrams $k_BT/(-W_Q)$~vs.~$n$ for \mbox{$W_1>0$} and several values of $U/(-W_Q)$ and \mbox{$k=z_2W_2/z_1W_1$}: (a)~\mbox{$U/(-W_Q)=0.2$}, \mbox{$k=-0.2$}; (b)~\mbox{$U/(-W_Q)=0.2$}, \mbox{$k=-0.4$}; (c)~\mbox{$U/(-W_Q)=0.6$}, \mbox{$k=-0.2$}; and (d)~\mbox{$U/(-W_Q)=0.6$}, \mbox{$k=-1$}.  Dotted, solid and dashed lines indicate first order, second order and ``third order'' boundaries, respectively.}
        \label{rys:U02U06}
\end{figure*}

For the on-site interaction \mbox{$U/(-W_Q)<(2/3)\ln2$} and \mbox{$U/(-W_Q)>1$} only the second order \mbox{CO--NO} transitions occur with increasing temperature.
For \mbox{$(2/3)\ln2<U/(-W_Q)<0.62$} the first order \mbox{CO--NO} transition  appears near \mbox{$n=1$} with a~\emph{tricritical point} $TC$ connected with a change of transition order (e.~g. figure~\ref{rys:U02U06}c). The $TC$ for \mbox{$n=1$} is located at \mbox{$k_BT/(-W_Q)=1/3$} and \mbox{$U/(-W_Q)=2/3\ln2$}.
In the range \mbox{$0.62<U/(-W_Q)<1$} the first order \mbox{CO--CO} line appears, which is ended at an~\emph{isolated critical point} $IC$ of the liquid-gas type (cf.~figure~\ref{rys:U08U094}a). In this case we have also a~\emph{critical end point} $CE$, where three boundary lines (one of second order: \mbox{CO--NO} and two of first order: \mbox{CO--NO} and \mbox{CO--CO}) connect together. The two CO phases: {\HCO} and {\LCO} are distinguishable only in the neighbourhood of the first order line \mbox{CO--CO} (\mbox{\HCO--\LCO}). This line is associated with the \mbox{\HCOB--\LCOB} transition in GS. The first order transition \mbox{CO--NO}  can exist inside the region \mbox{$0.79<n<1$} and \mbox{$0.5<n<1$}. The lines consisting of $TC$, $IC$ and $CE$ meet at a \emph{new multicritical point}, which coordinates are \mbox{$n=0.79$}, \mbox{$U/(-W_Q)=0.62$} and \mbox{$k_BT/(-W_Q)=0.24$}, approximately.

The phase diagrams obtained for attractive \mbox{$W_2<0$} are essentially different from those for \mbox{$W_2>0$}.
The main difference is that at sufficiently low temperatures PS states are stable. In the ranges of PS states occurrence the homogeneous phases can be metastable (if \mbox{$\partial \mu/ \partial n>0$}) or unstable (if \mbox{$\partial \mu/ \partial n<0$}). In the homogeneous phases occurring at higher temperatures (above the regions of PS occurrence) the stability condition \mbox{$\partial\mu/\partial n > 0$} is fulfilled.

For clarity of the presentation we will discuss the behaviour of the system at \mbox{$T>0$} for \mbox{$W_2<0$} distinguishing three regimes of $U/(-W_Q)$: on-site attraction (section~\ref{subsec:FTW2att0Uatt}), strong on-site repulsion (section~\ref{subsec:FTW2att0Urepstrong}) and weak on-site repulsion (section~\ref{subsec:FTW2att0Urepsmall}).

We also distinguish three different {\PSNDef} states and four different {\PSCDef} states (labeled by subscripts $A$, $B$, $C$ or superscript $*$). All states in a~particular group are states with the same type of phase separation (i.~e. CO--NO or CO--CO), however they occur in different regions of the phase diagram and such distinction has been introduced to clarify the presented diagrams (cf. especially figure \ref{rys:fixedn}). A~similar distinction has been done for all critical points connected with the phase separation (four $B$-type points: $B$, $B'$, $B''$, $B^*$, three $T$-type points: $T$, $T'$, $T''$ and several points of $H$-, $E$-, $F$- types). The lines \mbox{$M$-$N$-$O$} and \mbox{$X$-$Y$} indicate the first order transitions between PS states on the $k_BT/(-W_Q)$ vs.~$n$ diagrams (cf. figures \ref{rys:U02U06} and~\ref{rys:U08U094}).

\subsection{The case of on-site attraction}\label{subsec:FTW2att0Uatt}

For any on-site attraction (\mbox{$U\leq0$}) the phase diagrams are qualitatively similar, all (second order and ``third order'') transition temperatures decrease with increasing $U$ and for \mbox{$U=0$} the transition temperatures account for a~half of these in the limit \mbox{$U\rightarrow-\infty$}. In a~case of \mbox{$W_2<0$} beyond half-filling the PS states can be stable also at \mbox{$T>0$}.
Examples of the $k_BT/(-W_Q)$~vs.~$n$ phase diagrams evaluated for
\mbox{$U/(-W_Q)=0$}
and various ratios of \mbox{$k= z_2W_2/z_1W_1 \leq 0$} are shown in
figure~\ref{rys:U0}.
If \mbox{$0\leq|k|\leq1$} the (homogeneous) CO and NO phases are separated by the second order transition line.

When \mbox{$-0.6<k<0$} (figure~\ref{rys:U0}a) a~``third order'' transition takes place at low temperatures, leading first to the PS into two coexisting CO phases (\PSCA), while at still lower temperatures CO and NO phases coexist (\PSNA). The critical point for this phase separation
(denoted as $B$, we shall call this point a~\emph{bicritical endpoint}, BEP) is located inside the CO phase. The \mbox{$E$-$F$} solid line (we shall refer to \mbox{$E$-point} as a~\emph{critical endpoint}, CEP) is associated with continuous transition between two different PS states (\mbox{\PSNA--\PSCA}, the second order CO--NO transition occurs in the domain with lower concentration).

For \mbox{$k<-0.6$} (figure~\ref{rys:U0}b) the transition between PS states does not occur, the area of {\PSCA} stability vanishes and the critical point for the phase separation (denoted as $T$, which is a~\emph{tricritical point}, TCP) lies on the second order line \mbox{CO--NO}. As \mbox{$k \rightarrow -\infty$} the $T$-point occurs at \mbox{$n=1$} and the homogeneous CO phase does not exist beyond half-filling.

When \mbox{$k=-0.6$} the lower branch of the ``third order'' curve approaches the critical point ($H$) parabolically. The tricritical behaviour for \mbox{$k<-0.6$}  changes into the bicritical behaviour for \mbox{$k>-0.6$}. The $H$-point is a~\emph{higher order critical point} (HCP) and in this point the lines consisting of $B$ $E$, $F$, and $T$ points connect together (for fixed $U/(-W_Q)$). Similar scenario takes place also for the on-site repulsion apart from the bicritical behaviour connected with  the {\PSC} state, which can exist also for \mbox{$k>-0.6$}).

One should notice that a type of the critical point for separation (which can be BEP, TCP or HCP) is modified only by  a~change of the strength of the nnn attraction.  The location of the transition lines between homogeneous phases on the $k_BT/(-W_Q)$~vs.~$n$ diagrams is not affected by the value of $W_2$ (the transitions are at the same $k_BT/(-W_Q)$ as in the previous case of \mbox{$W_2>0$}, which only depends on the on-site interaction $U$ for fixed $n$). Effectively, the transition temperatures increase with increasing strength of attractive $W_2$.

The labels of the critical points for phase separations are given with a correspondence to those in~\cite{KC1975}, where the Ising model with nn and nnn interactions was considered. Notice that our model is equivalent to the Ising one in the \mbox{$U\rightarrow\pm\infty$} limits.

\subsection{The case of strong on-site repulsion}\label{subsec:FTW2att0Urepstrong}

For any \mbox{$U/(-W_Q)\geq 1$} the structure of the phase diagrams and the sequences of transitions are similar as those in the previous case (for corresponding values of $k$), but now the double occupancy of sites is strongly reduced due to repulsive $U$ and the phase diagrams are (almost) symmetric with respect to \mbox{$n=0.5$} (cf. figure~\ref{rys:GS}b and table~\ref{tab:table2}). $B'$, $H'$, $T'$, $E'$ and $F'$ points (as well as $B''$, $H''$, $T''$, $E''$ and $F''$ points) appear (cf.~\cite{KKR2010}), which correspond to $B$, $H$, $T$, $E$ and $F$ points, respectively. Critical behaviours at $A'$ and $A''$ points are the same as at $A$ points (\mbox{$A=B,H,T$}).

The exact symmetry occurs at \mbox{$U\rightarrow +\infty$}. In this limit the phase diagrams are
the diagrams for \mbox{$U\rightarrow-\infty$} with re-scaled axes: \mbox{$k_BT/(-W_Q)\rightarrow k_BT/(-4W_Q)$}
and \mbox{$n\rightarrow(1/2)n$}.

Transition temperatures are only weakly dependent on the on-site repulsion and  one finds a~small decrease of them with increasing $U$ for \mbox{$U/(-W_Q)\geq 1$}.

\subsection{The case of small on-site repulsion}\label{subsec:FTW2att0Urepsmall}

\begin{figure*}
        \centering
        \includegraphics[width=\figwidthone]{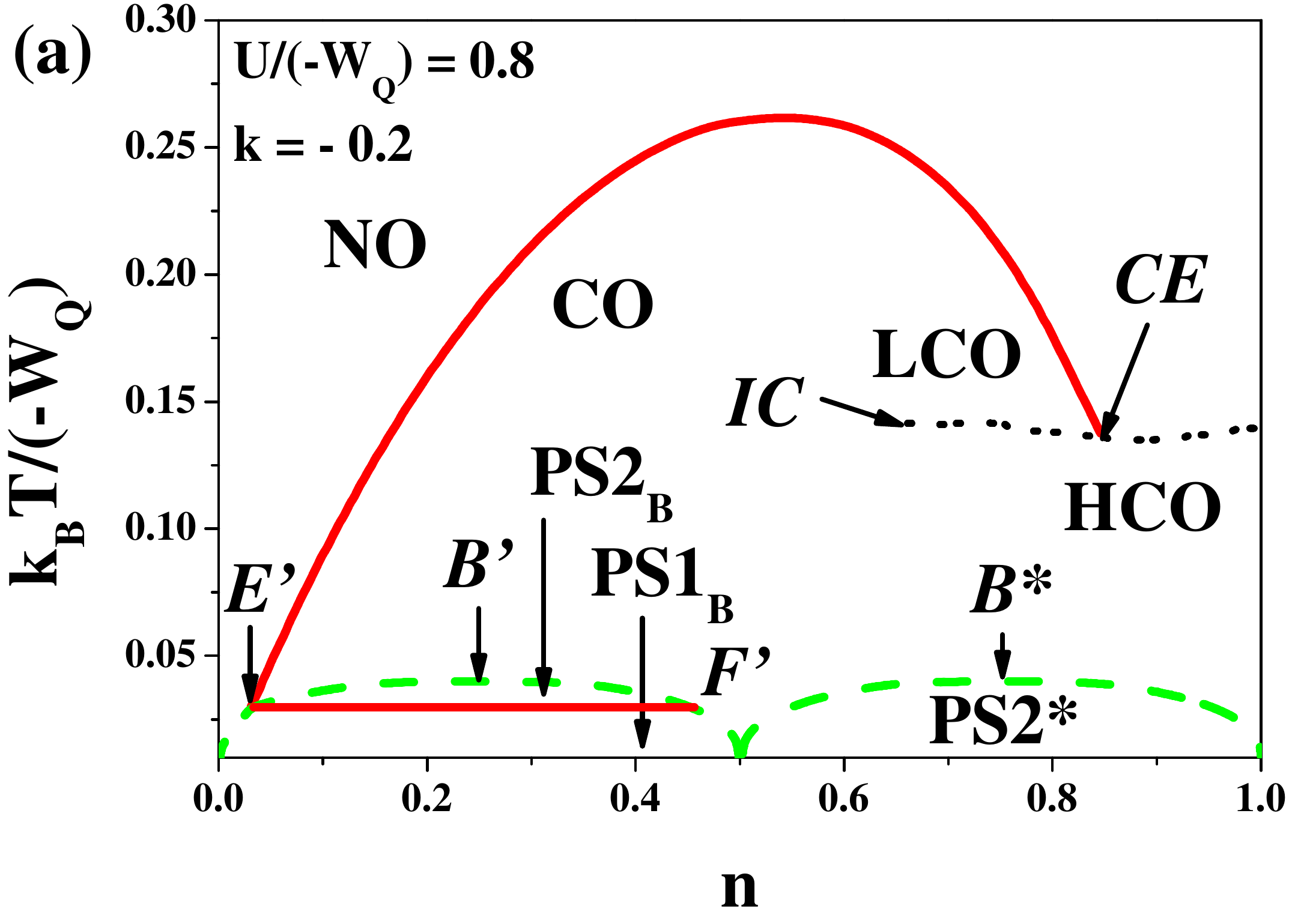}
        \includegraphics[width=\figwidthone]{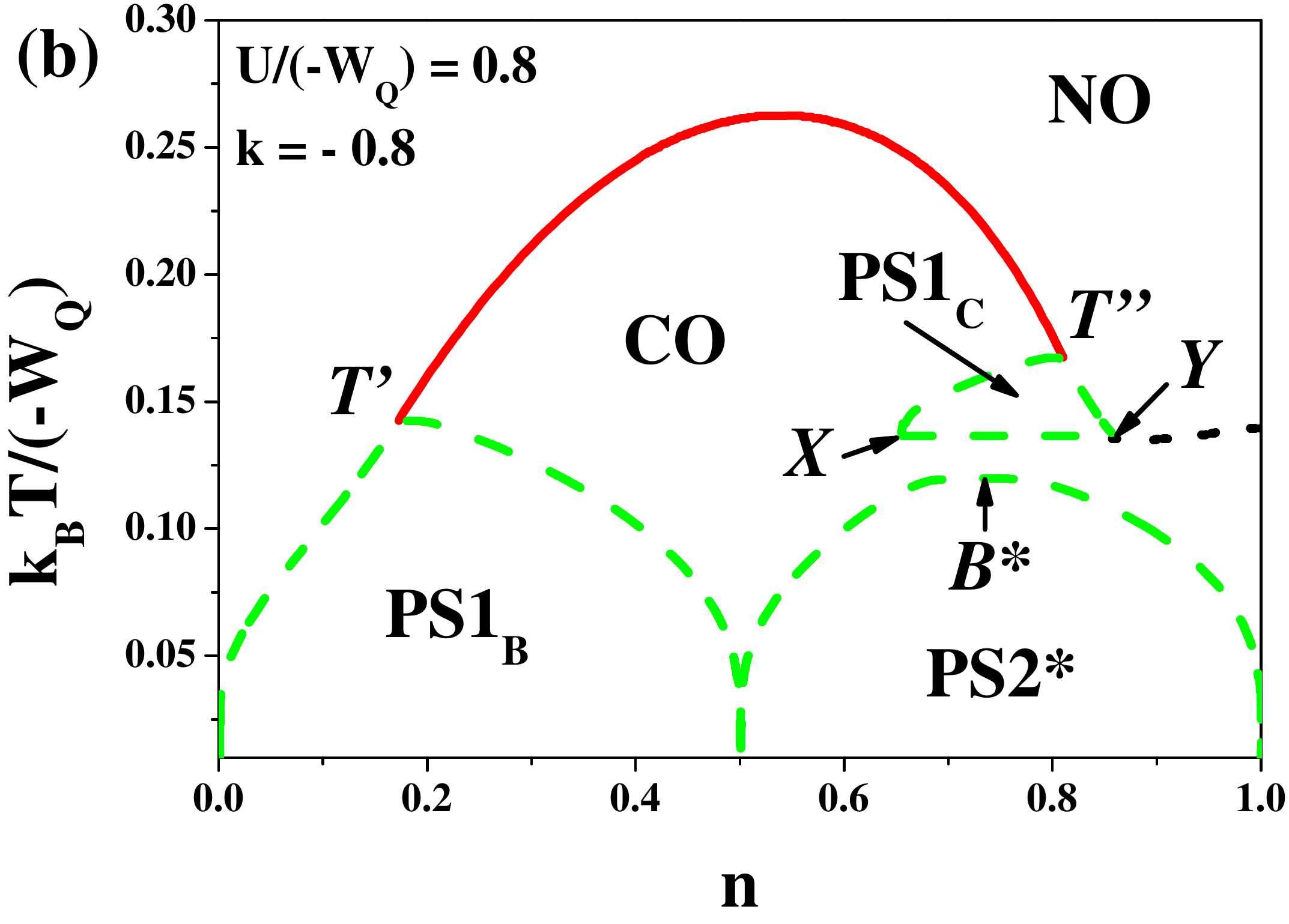}\\
        \includegraphics[width=\figwidthone]{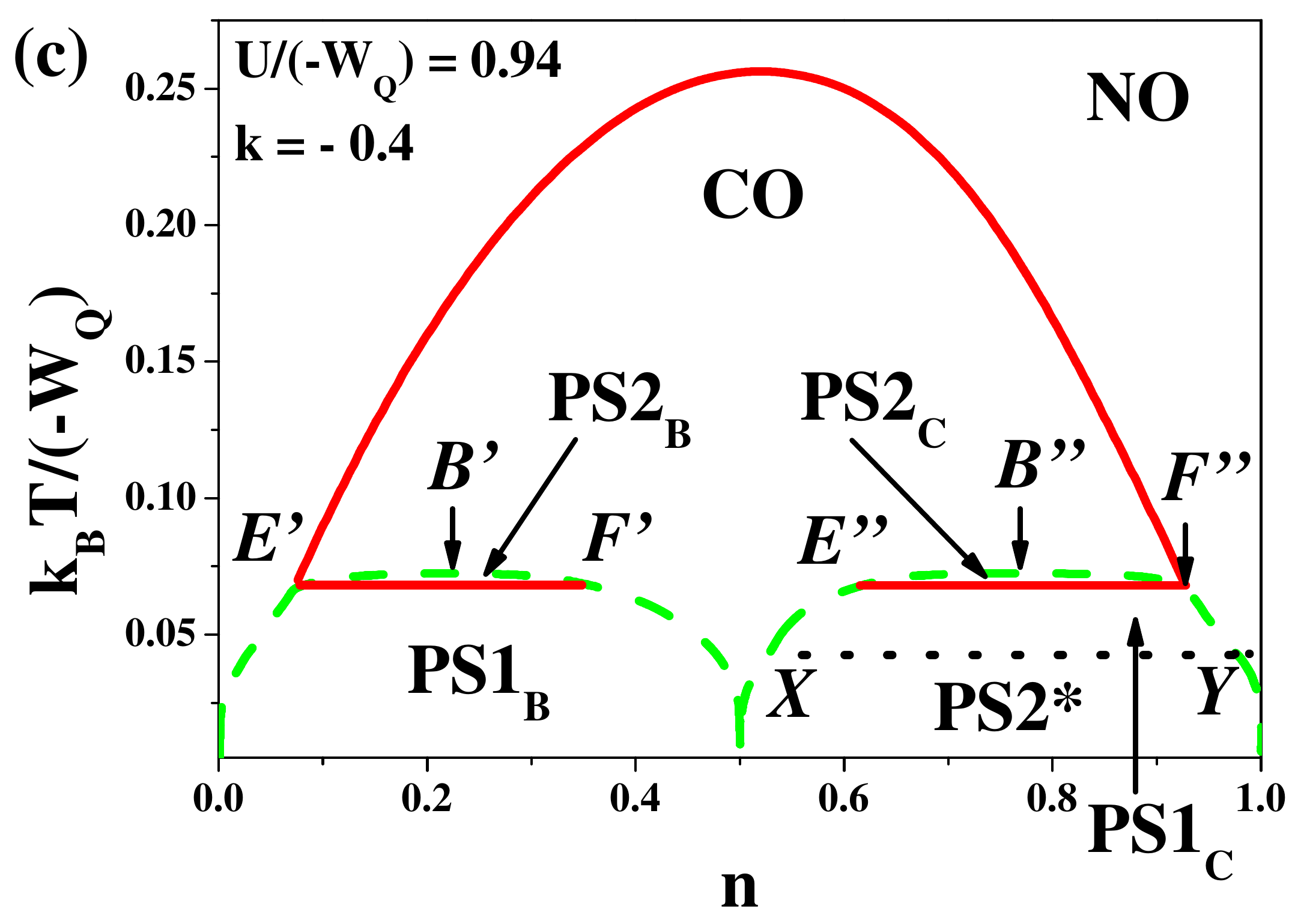}
        \includegraphics[width=\figwidthone]{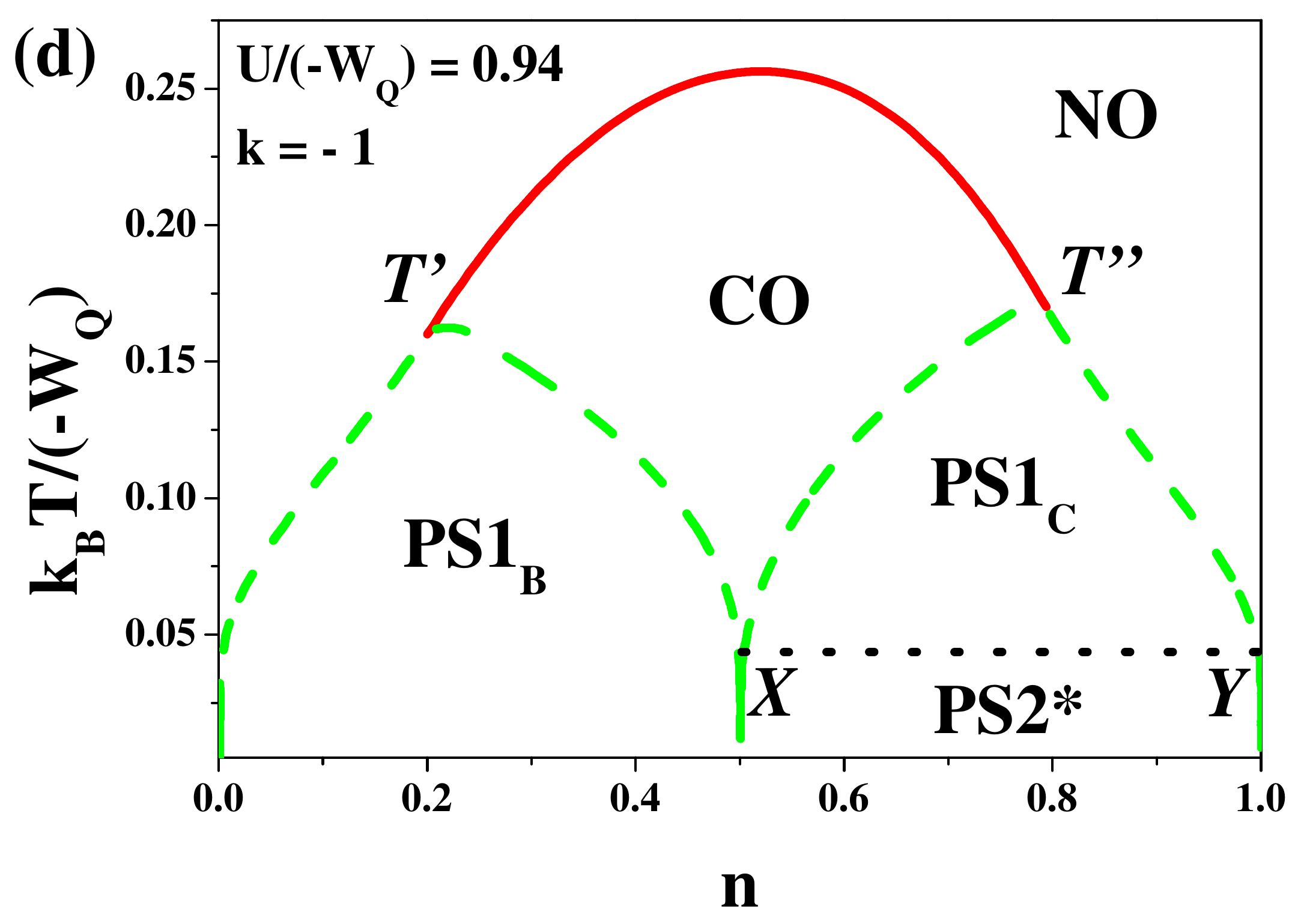}
        \caption{(Color online) Phase diagrams $k_BT/(-W_Q)$~vs.~$n$ for \mbox{$W_1>0$} and several values of $U/(-W_Q)$ and \mbox{$k=z_2W_2/z_1W_1$}: (a)~\mbox{$U/(-W_Q)=0.8$}, \mbox{$k=-0.2$}; (b)~\mbox{$U/(-W_Q)=0.8$}, \mbox{$k=-0.8$}; (c)~\mbox{$U/(-W_Q)=0.94$}, \mbox{$k=-0.4$}; and (d)~\mbox{$U/(-W_Q)=0.94$}, \mbox{$k=-1$}.  Dotted, solid and dashed lines indicate first order, second order and ``third order'' boundaries, respectively. Near \mbox{$n=1$}, on the right of the {\PSC} occurrence region, the CO phase is stable, what is not shown on the diagrams (c) and (d) explicitly.}
        \label{rys:U08U094}
\end{figure*}
\begin{figure*}
        \centering
        \includegraphics[width=\figwidthone]{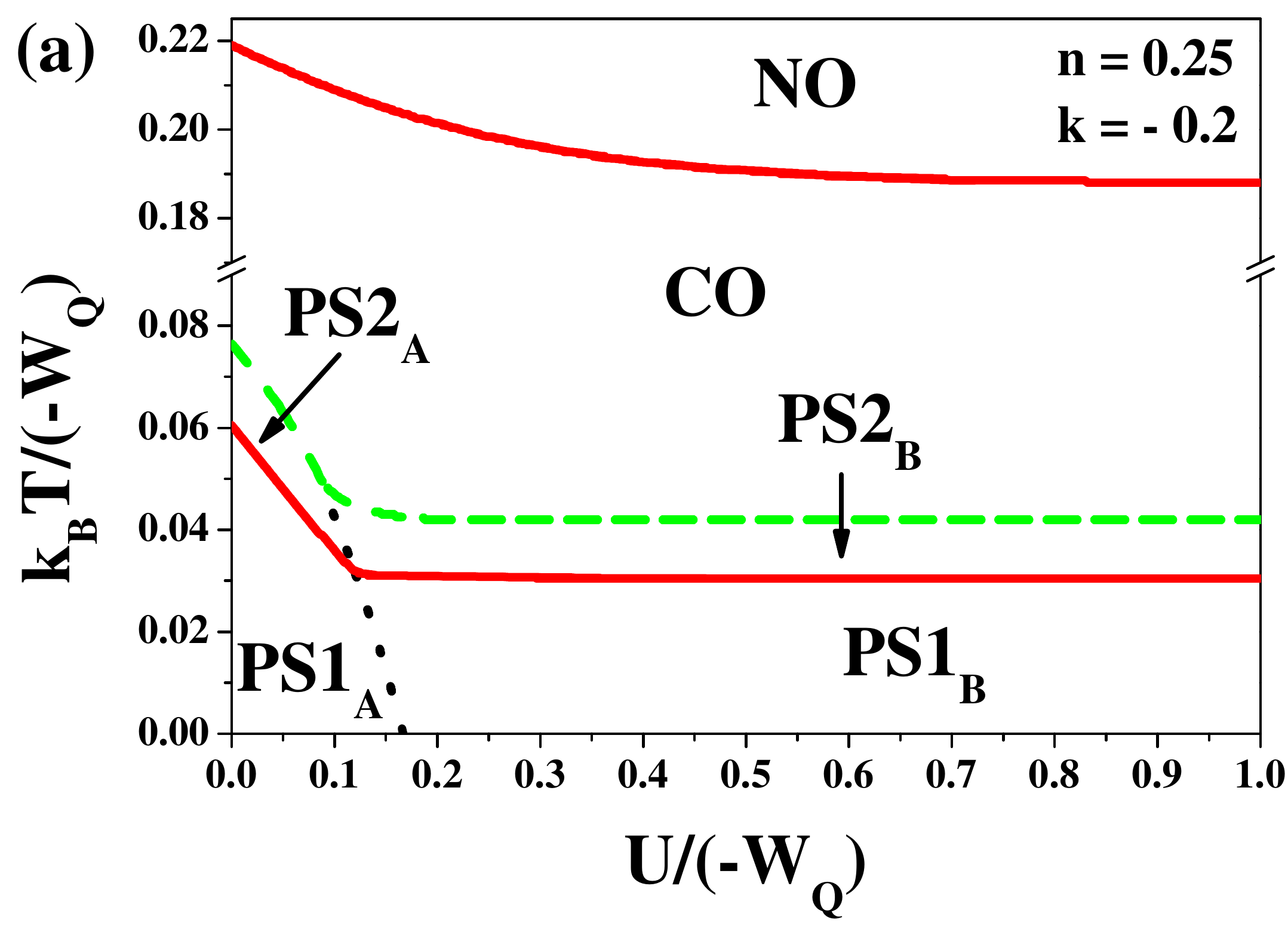}
        \includegraphics[width=\figwidthone]{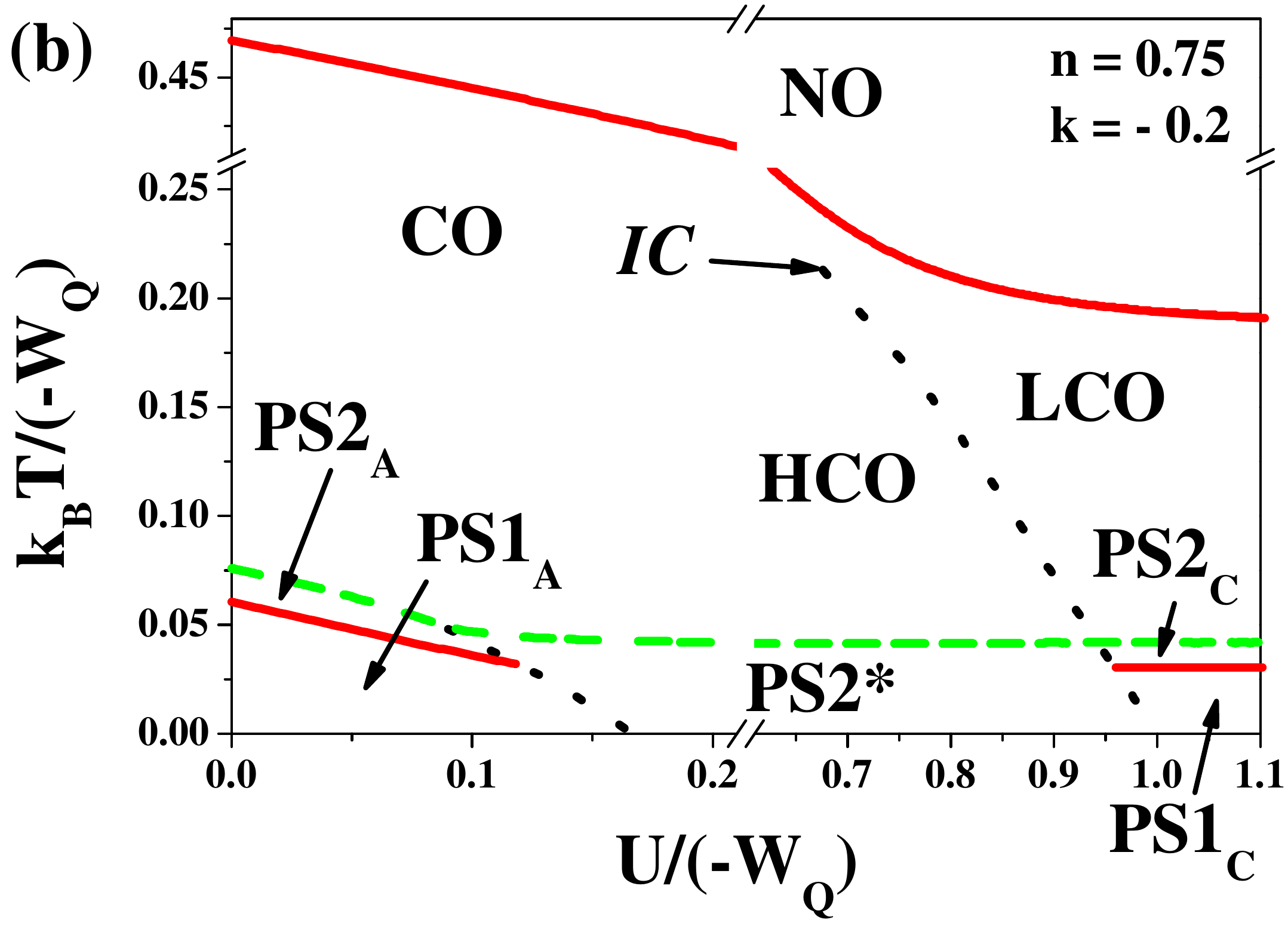}\\
 		\includegraphics[width=\figwidthone]{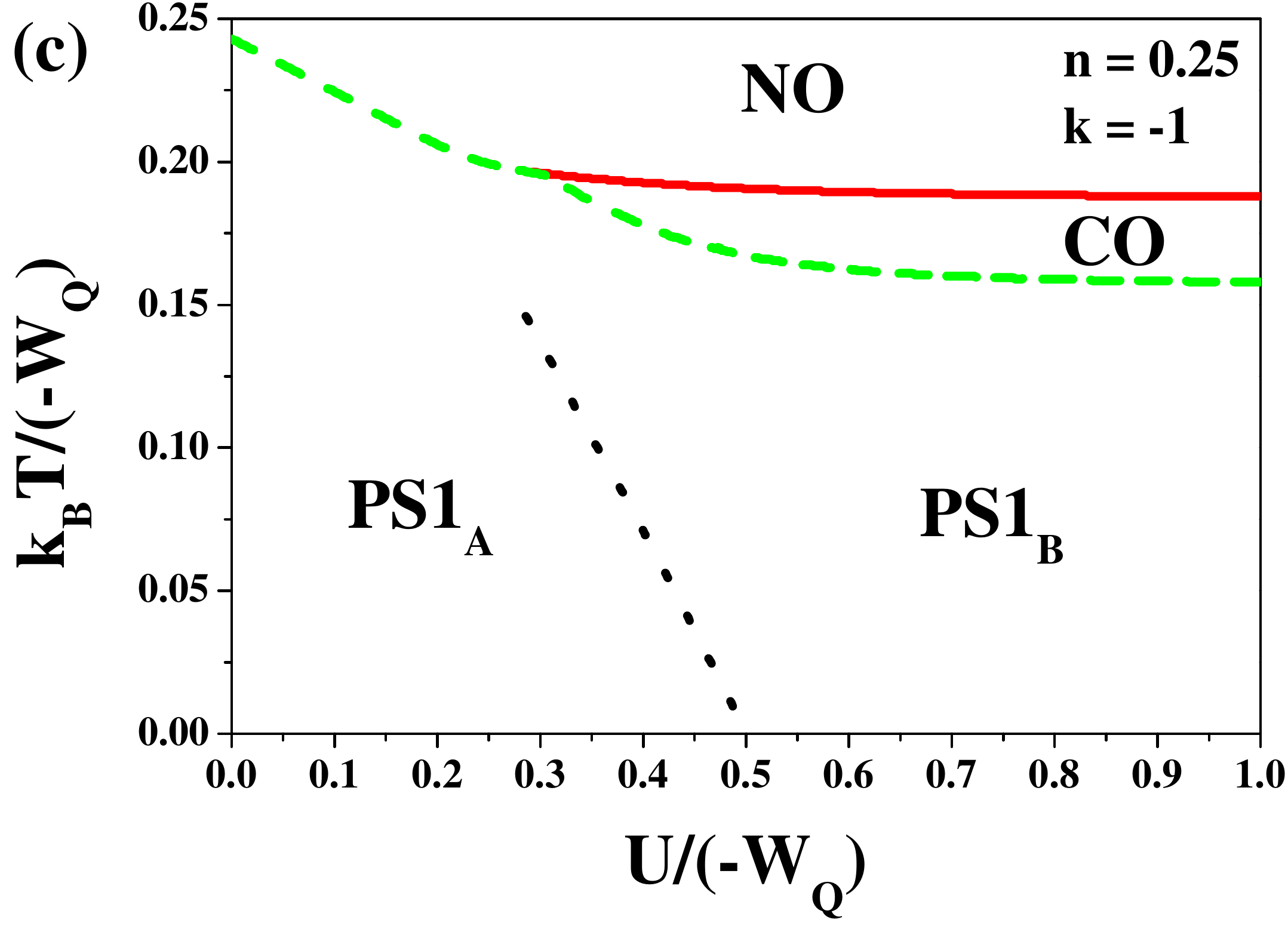}
        \includegraphics[width=\figwidthone]{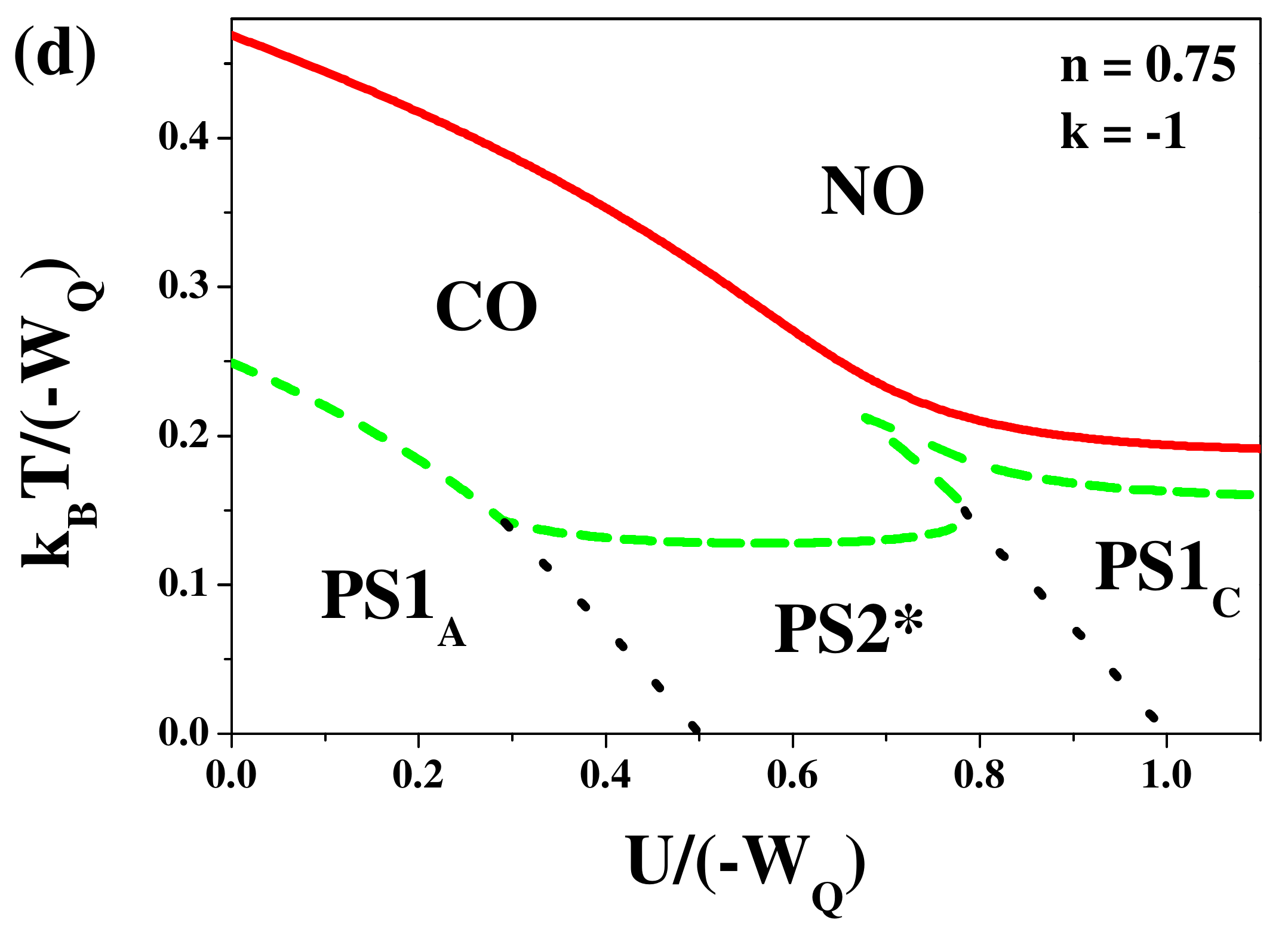}
        \caption{(Color online) Phase diagrams $k_BT/(-W_Q)$~vs. $U/(-W_Q)$ for several values of $n$ and \mbox{$k=z_2W_2/z_1W_1$}: (a)~\mbox{$k=-0.2$}, \mbox{$n=0.25$}; (b)~\mbox{$k=-0.2$}, \mbox{$n=0.75$}; (c)~\mbox{$k=-1$}, \mbox{$n=0.25$}; and (d)~\mbox{$k=-1$}, \mbox{$n=0.25$}. Dotted, solid and dashed lines indicate first order, second order and ``third order'' boundaries, respectively. One should notice that some axes are broken.}
        \label{rys:fixedn}
\end{figure*}

The range \mbox{$0<U/(-W_Q)<1$} is the most interesting one and the phase diagrams  are more complicated than those in previous cases. Due to the variety of the behaviour in this regime of on-site interaction we only present some particular examples of the phase diagrams (figures~\ref{rys:U02U06} and~\ref{rys:U08U094}).

In this range of $U/(-W_Q)$ an occurrence of the {\PSC} state for \mbox{$0.5<n_-<n<n_+<1$} is possible and the critical point for this phase separation (BEP, denoted as $B^*$) is located inside the CO phase also for \mbox{$|k|\geq0.6$}. It contrasts with BEPs mentioned previously (i.~e. $B$, $B'$ and $B''$), which occur only for \mbox{$|k|<0.6$}. The region of {\PSC} occurrence extends from the ground state if \mbox{$U/|z_2W_2|>1$}.

When \mbox{$0<U/(-W_Q)<2/3\ln2$} one can see on the phase diagrams
a~new behaviour, i.~e. discontinuous transitions between PS states connected with the \mbox{$M$-$N$-$O$} line. On figure \ref{rys:U02U06}b two such transitions are presented, i.~e. {\PSNA}--{\PSNB} and {\PSNA}--{\PSC}. There are also possible {\PSCA}--{\PSCB} and {\PSCA}--{\PSC} transitions, which are not shown there. All transitions between homogeneous phases are second order in this range.

In the range \mbox{$2/3\ln2<U/(-W_Q)<1$}, besides the behaviours mentioned previously, the critical points connected with transitions between homogeneous phases can appear on the phase diagrams:
(i)~$TC$ for \mbox{$U/(-W_Q)<0.62$} and the first order \mbox{CO--NO} boundary near \mbox{$n=1$} (cf.~figures~\ref{rys:U02U06}c and \ref{rys:U02U06}d) and
(ii)~$CE$ and $IC$ for \mbox{$U/(-W_Q)>0.62$} with the first order \mbox{\HCO--\LCO} transition slightly dependent on the electron concentration (cf.~figures~\ref{rys:U08U094}).
It leads to an appearance of isolated areas of {\PSNC} state stability (only for \mbox{$k\leq-0.6$}, figure~\ref{rys:U08U094}b) and to discontinuous transitions between two PS states:
(i)~\mbox{\PSC--\PSNC} (figures~\ref{rys:U08U094}c and \ref{rys:U08U094}d) and
(ii)~\mbox{\PSC--\PSCC} (only for \mbox{$|k|<0.6$}, cf. also figure~\ref{rys:fixedn}b) in a~restricted ranges of $U/(-W_Q)$.
These behaviours are connected with the \mbox{$X$-$Y$} line, which is linked with the first order border lines \mbox{CO--NO} and \mbox{CO--CO}. For \mbox{$0>k>-0.6$} above the \mbox{$X$-$Y$} line a~second order \mbox{\PSNC--\PSCC} transition also occurs. The temperature associated with \mbox{$X$-$Y$} line is independent of $k$ and depends only on $U/(-W_Q)$.

The finite temperature phase diagrams as a~function of $U/(-W_Q)$ at fixed $n$  for \mbox{$k=-0.2$} and \mbox{$k=-1$} are shown in figure~\ref{rys:fixedn}. The first order boundaries: (i)~\mbox{\PSNA--\PSNB} and \mbox{\PSCA--\PSCB} (on the diagrams for \mbox{$n=0.25$}) and (ii)~\mbox{\PSNA--\PSC} and \mbox{\PSCA--\PSC} (on the diagrams for \mbox{$n=0.75$}) are associated with the \mbox{$M$-$N$-$O$} line, whereas the first order boundaries: (iii) \mbox{\PSC--\PSNC} and \mbox{\PSC--\PSCC} (on the diagrams for \mbox{$n=0.75$}) are connected with the \mbox{$X$-$Y$} lines. For \mbox{$k=-0.2$} and \mbox{$n=0.25$} at higher temperatures the {\PSNA} and the {\PSNB} are not distinguishable and on the diagram (figure~\ref{rys:fixedn}c) the first order boundary line ends at a~critical point of the liquid-gas type (similar to $IC$). In figure~\ref{rys:fixedn}d for \mbox{$n=0.75$} one can also see following sequence of transitions with increasing temperature: \mbox{\PSC--CO--\PSNC--CO--NO}. It is interesting to notice that the {\PSNC} state exists here at higher temperatures than the homogeneous CO phase (see also figure~\ref{rys:U08U094}b).

One should notice  that the ``third order'' boundaries in figure~\ref{rys:fixedn} are not the lines of BEPs (nor TCPs). For considered ranges of $U/(-W_Q)$ and $k$ the $n$-coordinates of these points fulfill the following conditions: (i)~$T$ and $B$ points: \mbox{$n<0.5$}, (ii)~$T'$ and $B'$ points: \mbox{$n<0.25$}, (iii)~$T''$, $B''$ and $B^*$ points: \mbox{$n>0.75$}. One should also remember that \mbox{$E$-$F$}, \mbox{$E'$-$F'$}, \mbox{$E''$-$F''$}, \mbox{$M$-$N$-$O$} and \mbox{$X$-$Y$} lines in their ranges of occurrence are independent of $n$.

\section{Charge-order parameter vs. temperature}\label{sec:termodynamics}

In this section we present two representative temperature dependencies of the charge-order parameter for fixed model parameters.

\begin{figure}
        \centering
        \includegraphics[width=\figwidth]{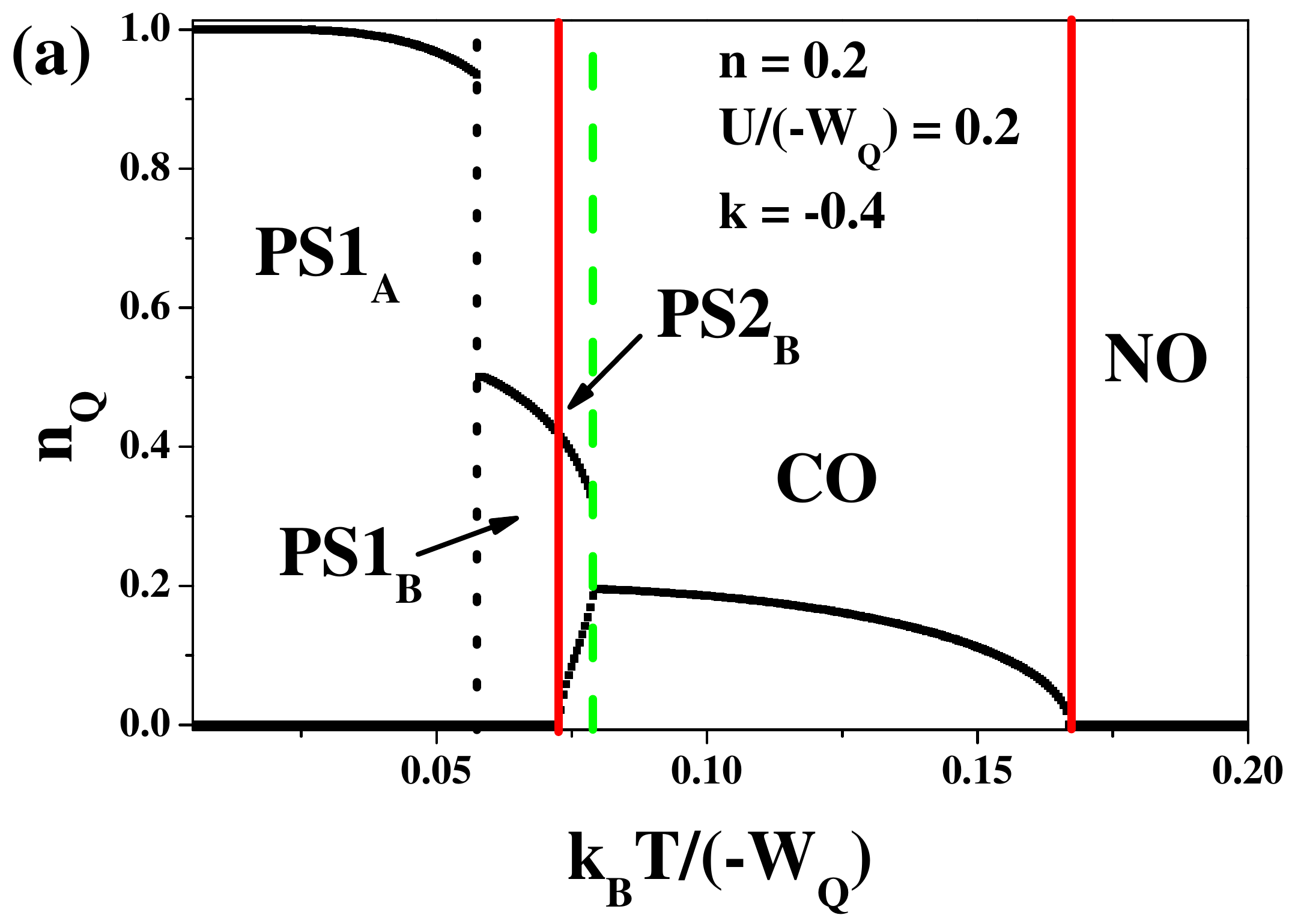}
        \includegraphics[width=\figwidth]{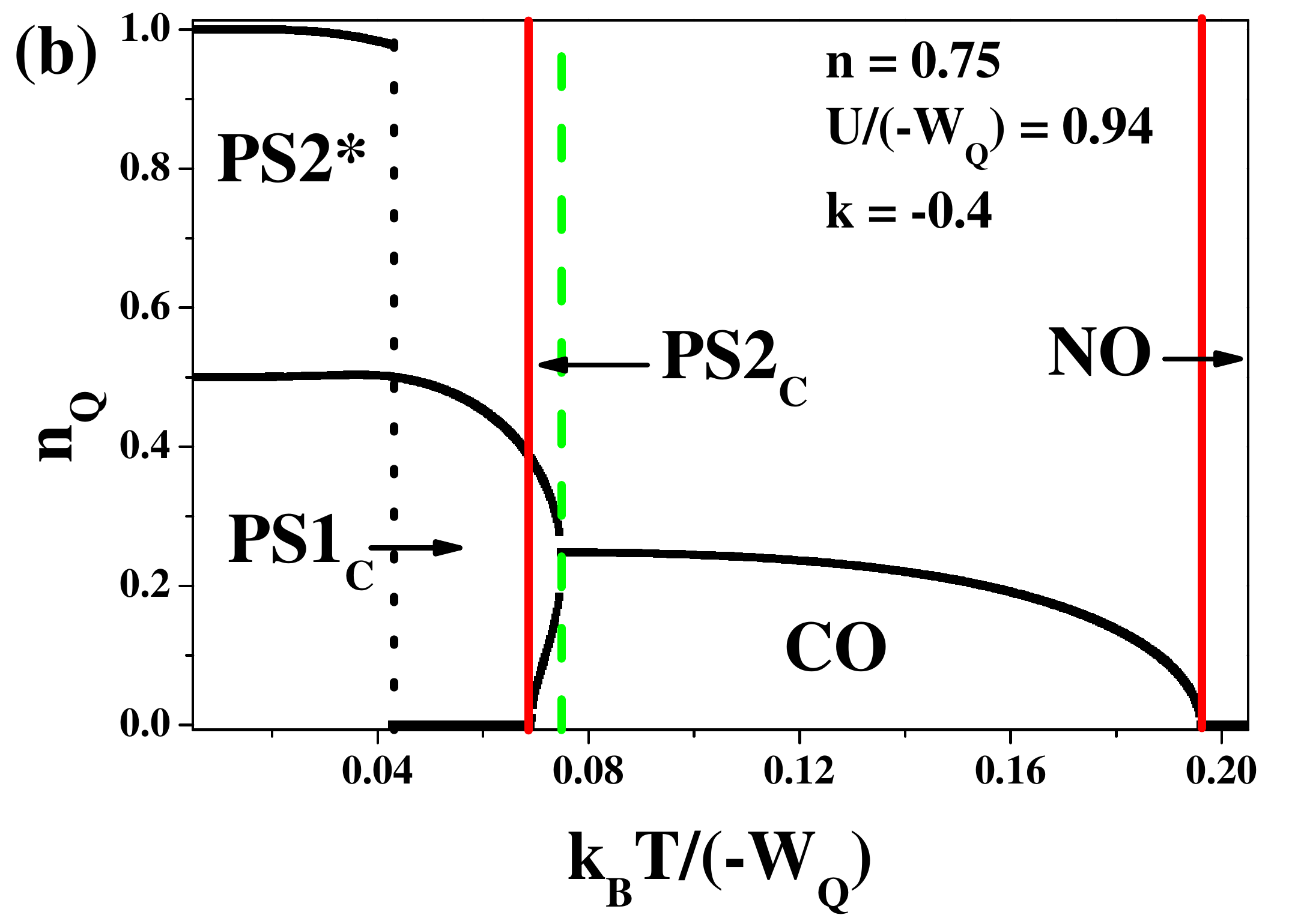}
        \caption{(Color online) Temperature dependencies of the charge-order parameter $n_Q$ for (a) \mbox{$U/(-W_Q)=0.2$}, \mbox{$n=0.2$} and \mbox{$k=-0.4$}; and (b) \mbox{$U/(-W_Q)=0.94$}, \mbox{$n=0.75$} and \mbox{$k=-0.4$}.}
        \label{rys:parnQ}
\end{figure}

In figure~\ref{rys:parnQ}a we have plotted the charge-order parameter $n_Q$ as a function of $k_BT/(-W_Q)$ for \mbox{$U/(-W_Q)=0.2$}, \mbox{$n=0.2$} and \mbox{$k=-0.4$}. At \mbox{$k_BT/(-W_Q)=0.058$} one observes a~discontinuous change of \mbox{$n_Q>0$} in one domain (in the other \mbox{$n_Q=0$}), connected with the first-order transition \mbox{\PSNA--\PSNB}. For \mbox{$k_BT/(-W_Q)=0.073$} the continuous transition \mbox{\PSNB--\PSCB} occurs and $n_Q$ raises to non-zero value in the domain with lower electron concentration (cf. figure~\ref{rys:U02U06}b). At \mbox{$k_BT/(-W_Q)=0.079$} the domain with higher electron concentration vanishes continuously, a~``third order'' transition \mbox{\PSCB--CO} occurs and the whole system is characterized by one value of the charge order-parameter. Next, at \mbox{$k_BT/(-W_Q)=0.167$} $n_Q$ goes to zero, i.~e. one has the second order \mbox{CO--NO} transition.

Finally, let us comment on the temperature dependence of $n_Q$ for \mbox{$U/(-W_Q)=0.94$}, \mbox{$n=0.75$} and \mbox{$k=-0.4$} (figure~\ref{rys:parnQ}b).
At \mbox{$k_BT/(-W_Q)=0.043$} one observes a~first order \mbox{\PSC--\PSNC} transition, which is connected with discontinuous change of $n_Q$ in the domain with higher electron concentration (cf.~figure~\ref{rys:U08U094}c). Next, at \mbox{$k_BT/(-W_Q)=0.069$} a~continuous \mbox{\PSNC--\PSCC} transition occurs (now one has continuous change of $n_Q$ in the domain with higher electron concentration). At higher temperatures one can notice a~``third order'' \mbox{\PSCC--CO} transition (at \mbox{$k_BT/(-W_Q)=0.075$})  and a~second order \mbox{CO--NO} transition (at \mbox{$k_BT/(-W_Q)=0.196$}).

\section{Beyond the two-sublattice orderings for repulsive $W_2$}\label{sec:beyond}

The nn repulsion \mbox{$W_1>0$}, as well as the nnn attraction \mbox{$W_2<0$}, favour two-sublattice ordering and in such a~case no other types of long-range order can occur. On the other hand, the repulsive \mbox{$W_2>0$} compete with \mbox{$W_1>0$} reducing stability of two-sublattice orderings and can yield the appearance of multi-sublattice orderings. In this section we consider charge-orderings on the regular lattices taking into account not only two-sublattice orderings. We will consider them as follows.

More general, we can define the charge-order parameter as \mbox{$n_{\vec{q}}=\sum_i n_i\exp{(i\vec{q}\cdot\vec{R}_i/a)}$}, where $\vec{R}_i$ determine a~location of $i$ site in the space and $a$ is a hypercubic lattice constant.

The four-sublattice orderings can be considered on the alternate lattices, in which both interpenetrating sublattices are also alternate lattices. Examples of such lattices are 1D-chain and 2D-square (SQ) lattice.

When we consider four-sublattice orderings in the GS, the homogeneous phases not mentioned in section~\ref{sec:GS} are found to occur for \mbox{$W_2>0$} in some definite ranges of \mbox{$k$} and \mbox{$U/z_1W_1$}.

\begin{figure}
        \centering
        \includegraphics[width=\figwidth]{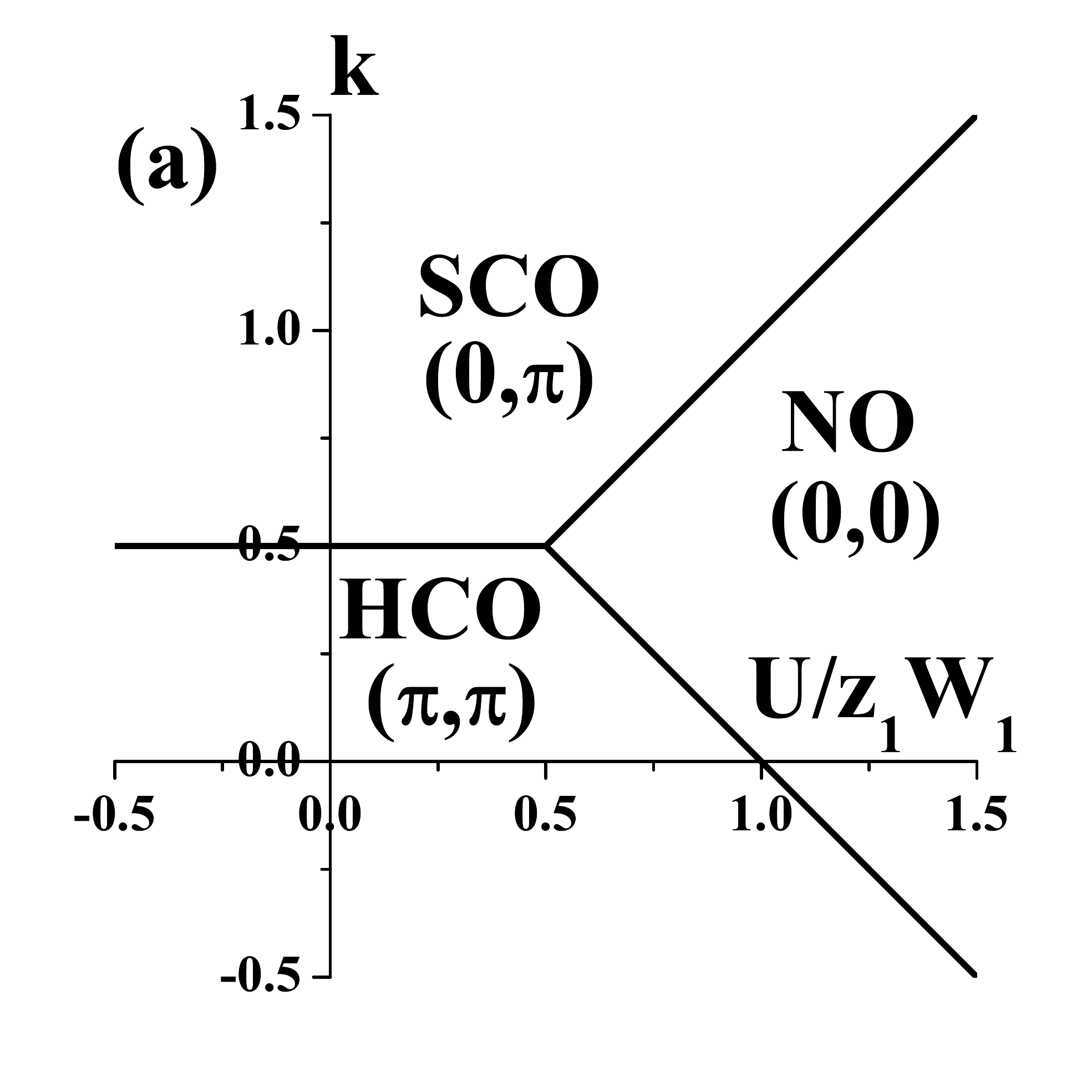}
        \includegraphics[width=\figwidth]{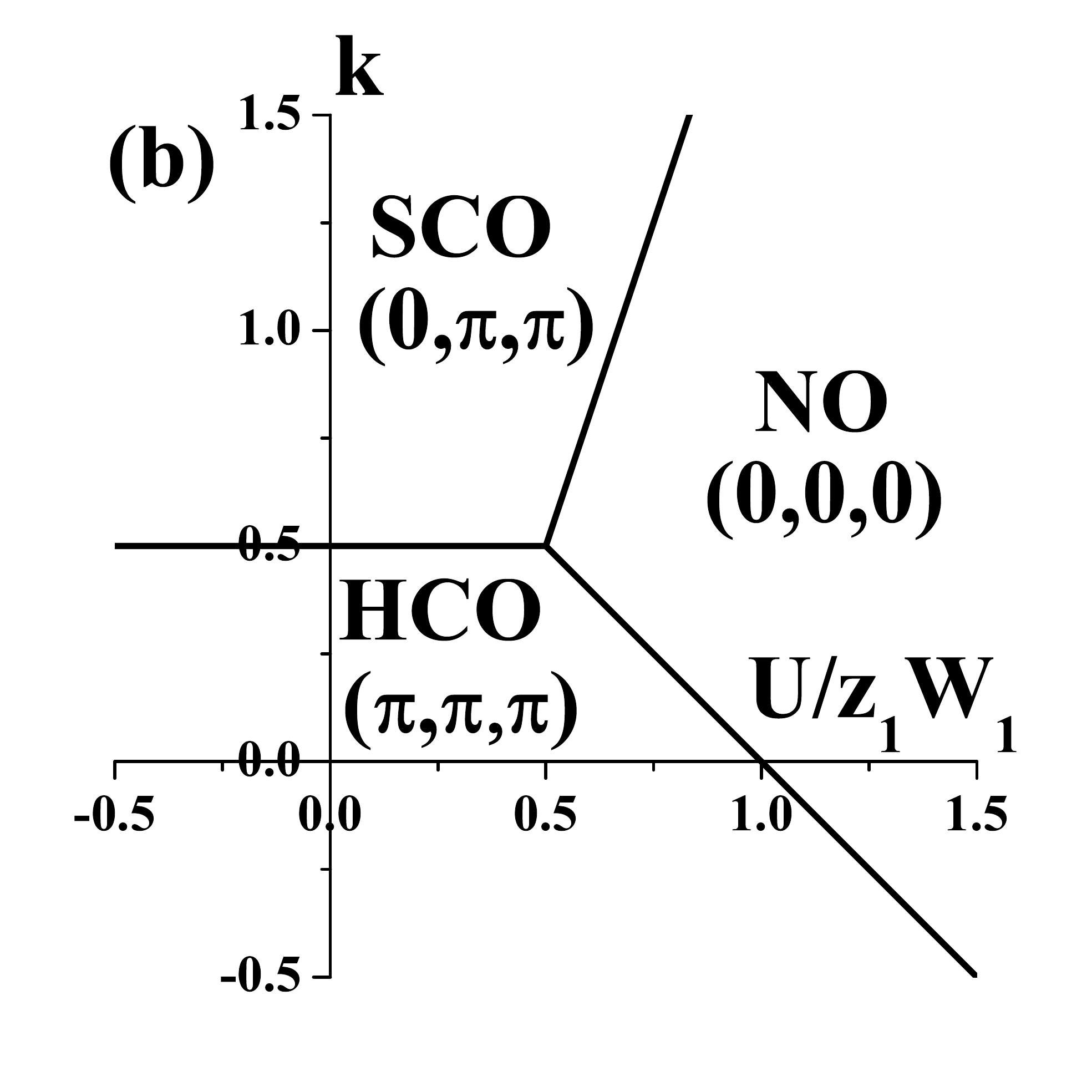}
        \caption{Ground state phase diagrams for \mbox{$n=1$} with consideration of the multi-sublattice orderings: (a)~for 2D-square lattice (for 1D-chain the region of SCO is replaced by that of ICO), (b)~for SC lattice in \mbox{$d=3$}. The phases are also labeled by vector $\vec{q}$ (\mbox{$k=z_2W_2/z_1W_1$}, \mbox{$W_1>0$}).}
        \label{rys:GSfour}
\end{figure}

For example at half-filling a~so-called \emph{island} charge ordered phase (ICO, \ldots2200\ldots, in \mbox{$d=1$}) or \emph{stripe} charge ordered phase (SCO, in \mbox{$d=2$}, \mbox{$\vec{q}=(0,\pi)$})
can occur for \mbox{$z_2W_2/z_1W_1>0.5$}.
In HCO phase we have \mbox{$\vec{q}=\pi$} in \mbox{$d=1$} and
\mbox{$\vec{q}=(\pi,\pi)$} in \mbox{$d=2$}.
The GS phase diagram taking into account the four-sublattice orderings on 2D-square lattice for \mbox{$n=1$}, \mbox{$W_1>0$} is shown in figure~\ref{rys:GSfour}a. In the case of 1D-chain the range of ICO occurrence is the same as that of SCO in 2D.  In both cases the VA results (for \mbox{$n=1$}) are consistent with exact results~\cite{J1994,BJK1996,FRU2001,RK1982}.

On the SC lattice we cannot consider four-sublattice orderings, because the two interpenetrating sublattices are fcc sublattices and they are not alternate lattices. In case of such a~lattice at half-filling the following three types of commensurate charge orderings should be considered: (i)~\mbox{$\vec{q}=(0,0,\pi)$} (\emph{plane} charge ordered phase, PCO), (ii)~\mbox{$\vec{q}=(0,\pi,\pi)$} (SCO) and (iii)~\mbox{$\vec{q}=(\pi,\pi,\pi)$} (HCO).
The GS phase diagram for SC lattice and \mbox{$n=1$} is shown  in figure~\ref{rys:GSfour}b. The PCO phase does not occur in GS and a~region of the NO phase stability is extended in comparison to the lower dimension cases.

In all CO phases mentioned previously the number of electrons on the particular site can be \mbox{$n_i=0$} or \mbox{$n_i=2$} and the charge order parameter in each phase has a~maximum possible value, i.~e.~\mbox{$n_{\vec{q}}=1$}. In the NO \mbox{$n_{\vec{q}}=0$} (\mbox{$n_i=1$} at every site).

One should notice that a~region of the HCO phase occurrence (on the $k$~vs.~$U/(z_1W_1)$ diagram) does not depend on the lattice dimension. This result is in agreement with GS phase diagrams obtained in section~\ref{sec:GS}. The discontinuous transition \mbox{HCO--NO} is at \mbox{$k=1-U/z_1W_1$} what is equivalent to \mbox{$U/(-W_Q)=1$} (for \mbox{$k<0.5$} and \mbox{$n=1$}).

Let us stress that we have not analyzed the four-sublattice orderings at \mbox{$T>0$}. They can be stable at sufficiently low temperatures, such as SCO near half-filling for \mbox{$k\gtrsim 0.5$}~\cite{KHF2008} or some states with phase separation between different CO phases or between CO and NO phases (for \mbox{$n\neq1$})~\cite{BL1980}. Thus, the finite temperature phase diagrams for \mbox{$W_2>0$} with taking into account the four-sublattice orderings can be in general more involved than those discussed in section~\ref{sec:FTW2}.

\section{Concluding remarks}\label{sec:conclusions}

In this paper we studied atomic limit of the extended Hubbard model with intersite nn repulsion $W_1$.
By taking into account for the first time the effects of nnn density-density interaction (attractive and repulsive \mbox{$W_2\lessgtr0$}) and including into consideration the states with phase separation (involving CO) our paper substantially extends and generalizes the results of previous works concerning the model considered \cite{R1979,MRC1984}.
Let us summarize the most important conclusions of our work:

(i) Depending on the values of the interaction parameters and the electron concentration,  the system can exhibits not only several charge ordered and nonordered homogeneous phases, but also (for attractive $W_2$) various phase separated states involving charge orderings: \PSNDef (CO-NO) and \PSCDef (CO-CO).

(ii) Obtained phase diagrams have a~very rich structure with multicritical behaviours.
The regions of PS states (both {\PSNDef} and {\PSCDef}) stability expand with increasing of the next-nearest-neighbour attraction. Moreover, the transitions between PS states can be both continuous and discontinuous  such as those between homogeneous phases.

(iii) The value of \mbox{$W_2<0$} determines a type of the multicritical point associated with PS states (which we mentioned as \emph{bicritical}, \emph{tricritical} or \emph{high-order critical point}).

(iv) For repulsive \mbox{$W_2>0$} there is a~possibility of occurrence of multi-sublattice orderings, e.~g. stripe (or island) structures. In particular, these types of ordering have the lowest energy at \mbox{$T=0$} for \mbox{$n=1$} (cf. section~\ref{sec:beyond}) if \mbox{$k=z_2W_2/z_1W_1>0.5$} and \mbox{$U/z_1W_1>1$} as well as for \mbox{$n=0.5$} if \mbox{$k>0.5$} and \mbox{$U/z_1W_1\rightarrow+\infty$} and in such cases they can be stable also at sufficiently low temperatures.

Let us stress that for \mbox{$W_1>0$} and \mbox{$W_2<0$} the derived results are exact in the limit of infinite dimensions, where the MFA treatment of intersite interactions becomes the rigorous one.

Charge ordered phases are stable (for \mbox{$W_2=0$}) when the quantum perturbation by finite bandwidth is introduced \cite{FRU2001,LCC2000,D1996,BS1986,MRC1983}. Thus one can conclude that the {\PSNDef} and {\PSCDef} states, which involve the CO phases, should also occur in the presence of hopping term. The stability of the phase separated CO--NO state in finite temperature (for \mbox{$t_{ij}\neq0$} and \mbox{$W_2=0$}) in a~definite range of the electron concentration was confirmed by using dynamical mean field approximation~\cite{TSB2004}.
If \mbox{$t_{ij}\neq 0$} for \mbox{$U<0$} the on-site superconducting states can occur, whereas for \mbox{$U>0$} it is necessary to consider also magnetic orderings \mbox{\cite{MRR1990,MRC1983,RMC1981,RP1996,ITK2010}}. In such a~case various phase separation states involving superconducting and (or) magnetic orderings can also be stable.

For the case \mbox{$W_1<0$}, which was not analysed in the present work, the model (\ref{row:1}) (\mbox{$W_2=0$}) exhibits a phase separation \mbox{NO--NO} (electron droplet states) at low temperatures~\cite{BT1993,MM2010}. In this PS state different spatial non-ordered regions have different average electron concentrations. In such a~case, at higher temperatures only the homogeneous NO phase occurs.

The VA results  can be exact in the limit of infinite dimensions only. Below we discuss briefly some results obtained for \mbox{$W_2=0$} within other approaches.
In particular, the Bethe--Peierls--Weiss (BPW) treatment of the $W_1$ term predicts the following ranges of the existence of long-range CO at \mbox{$T=0$} for SQ and SC lattices in the limits: (a) \mbox{$U/z_1W_1 \rightarrow +\infty$} ({\LCO} phase): \mbox{$1/z_1<n<(2z_1-1)/z_1$} and (b) \mbox{$U/z_1W_1 \rightarrow - \infty$} ({\HCO} phase): \mbox{$2/z_1<n<2(z_1-1)/z_1$}. In particular:

(i)~\mbox{$d=2$} (SQ lattice):
    (a)~\mbox{$0.25<n<0.75$} and \mbox{$1.25<n<1.75$} ({\LCO}),
    (b)~\mbox{$0.5<n<1.5$} ({\HCO});

(ii)~\mbox{$d=3$} (SC lattice):
	(a)~\mbox{$0.17<n<0.83$} and \mbox{$1.17<n<1.83$} ({\LCO}),
	(b)~\mbox{$0.33<n<1.67$} ({\HCO});

(iii)~\mbox{$d=+\infty$} (hypercubic lattice):
    (a)~\mbox{$0<n<1$} and \mbox{$1<n<2$} ({\LCO}),
    (b)~\mbox{$0<n<2$} ({\HCO}).

The Monte Carlo calculations performed for SQ lattice \cite{P2006} yields for the case \mbox{$U\rightarrow + \infty$} even more restricted ranges for long-range CO ({\LCO}) at \mbox{$T=0$}: \mbox{$0.37<n<0.63$} and \mbox{$1.37 < n <1.63$}.
In this particular case the existence of percolations of the effective clusters has been confirmed. These percolations vanish at transition temperature.

The phase diagrams for \mbox{$W_1>0$} and \mbox{$W_2=0$} obtained by exact solution for the Bethe lattice \cite{MM2010} (which is equivalent to BPW approximation) have a~similar structure as the VA diagrams (even for small \mbox{$z_1=3$}). The main difference is a~reentrant behaviour found in the case of Bethe lattice for \mbox{$U<0$} if \mbox{$n<2/z_1$} and \mbox{$n>2(z_1-1)/z_1$}, where the sequence of phase transitions: \mbox{NO$\rightarrow$CO$\rightarrow$NO} can occur with increasing temperature. The transition temperatures determined in \cite{MM2010} are in general lower than those obtained in VA, but obviously in the limit of large coordination number the rigorous results for Bethe lattice reduce to those of VA.

Comparing the GS diagram obtained for \mbox{$W_2=0$} in VA (figure~\ref{rys:GS}a) with the exact one for 1D-chain~\cite{MM2008}, we notice that all the border lines are the same, although in the exact solution for \mbox{$d=1$} the long-range charge orderings in GS exist only in the ranges (i) \mbox{$0.5< n < 1$} and \mbox{$0 \leq U/(-W_Q) \leq 1 $}, (ii) \mbox{$n=0.5$} and \mbox{$U/(-W_Q)\geq0$}, (iii) \mbox{$n=1$} and \mbox{$U/(-W_Q)\leq1$} (what corresponds to the regions of \HCOB, {\LCO} and {\HCO} phases existence in figure~\ref{rys:GS}a, respectively). The values  of $\mu$ and $D$ obtained in VA are consistent with exact ones for arbitrary $n$ and $U/(-W_Q)$. Moreover, the GS phase diagrams as a function of $\mu$ for \mbox{$W_2=0$} derived within VA agree exactly with the corresponding rigorous solutions in \mbox{$d=1$} and \mbox{$d=2$}~\cite{BJK1996,FRU2001,PS1975,J1994}.

The above discussion implies that VA in the case \mbox{$W_2=0$} can give qualitatively reasonable results beyond the percolation thresholds also for lattices of finite dimensionality and this statement should also be true for \mbox{$W_2\neq0$}, at least for small attractive $W_2$.

The electron concentration $n$ and chemical potential $\mu$ are (thermodynamically) conjugated  variables in the bulk systems~\cite{MM2008}. However, one can fit the concentration rather than the potential in a~controlled way experimentally. In such a~case $\mu$ is a~dependent internal parameter, which is  determined by the temperature, the value of $n$, and other model parameters (cf.~(\ref{row:2})). Thus the obtained phase diagrams as a~function of the concentration are quite important  because in real systems $n$ can vary in a~large range and charge orderings are often found in extended ranges of electron doping (e.~g. in doped manganites~\cite{GL2003,FT2006,DHM2001,DKC2001,RAA2005,QBK2004}, nickelates~\cite{IFT1998,GSS1985} and the doped barium bismuthates~\cite{MRR1990,V1989,GK1994}).
In Bechgaard salts the concentration is \mbox{$n=1/2$}~\cite{IKM2004,SHF2004,MLC1990}. In charge transfer salts $n$ changes, dependent on the pressure, in the vicinity of \mbox{$n=2/3$}, whereas for several complex TCNQ salts $n$ is near \mbox{$n=1/2$} \cite{VHN2001,SHF2004,F2006,SS1981}. In cuprates (\cite{MRR1990,MR1997} and references therein) and in conducting polymers \cite{BCM1992} $n$ is near half-filling in the insulating state and it strongly changes under doping.

Although our model is (in many aspects) oversimplified, it can be useful in qualitative analysis of experimental data for real narrow-band materials and it can be used to understand better properties of several CO systems mentioned above and in section~\ref{sec:intro}.

In particular, our results predict existence of the phase separation (CO-NO, CO-CO) generated by the effective nnn attractive interactions and describe their possible evolutions and phase transitions with increasing $T$ and a change of $n$. The electron phase separation involving COs is shown experimentally in several systems quoted above, e.~g.~in \mbox{R$_{1-x}$Ca$_x$MnO$_3$} (\mbox{R$=$La, Bi, Nd, etc.}), at dopings ranging from \mbox{$x=0.33$} to \mbox{$x=0.82$} \cite{DHM2001,DKC2001,RAA2005,QBK2004,TSB2004}. Among the materials for which the on-site local electron pairing  (valence skipping) has been either established or suggested (cf. section~\ref{sec:intro}) the best candidates to exhibit the phase separation phenomena are the doped barium bismuthates (\mbox{BaPb$_{1-x}$Bi$_x$O$_3$} and \mbox{Ba$_{1-x}$K$_x$BiO$_3$})~\cite{V1989,GK1994,MRR1990,RP1996}. For these systems, being oxide perovskites, a~very large dielectric constant strongly weakens the long-range Coulomb repulsion, which is the main factor preventing the phase separation~\cite{EK1993}.

Our results show that also the transitions at \mbox{$T>0$} between various homogeneous phases ({\HCO} and {\LCO}) and nonordered states can be either first order or continuous ones and both these types of the CO transitions are experimentally observed  in real narrow-band materials \cite{DHM2001,DKC2001,RAA2005,QBK2004}. Moreover, the theory predicts that with a~change of the model parameters ($U/(-W_Q)$, $W_2$, $n$) the system can exhibits various types of multicritical behaviour (including TCP, BCP, etc.) resulting from the competition of the on-site repulsion (\mbox{$U>0$})  and the effective intersite repulsion \mbox{$W_Q<0$}. In fact, some of charge ordered systems are found to exhibit the multicritical behaviour, e.~g. in organic conductor \mbox{(DI-DCNQI)$_2$Ag} (\mbox{$T_c=210$~K}) the temperature vs. pressure phase diagram shows continuous and first order boundaries with a~tricritical point~\cite{IKM2004}. The increasing pressure changes first the order of transition, resulting in a~tricritical point, then it yields a~complete suppression of charge orderings at any $T$.

\begin{acknowledgments}
The authors wish to thank R.~Micnas and T.~Kostyrko for helpful discussions and a~careful reading of the manuscript.
\end{acknowledgments}

\appendix*
\section{Site-dependent self-consistent VA equations}\label{app:equation}

Within the VA  the on-site interaction term is treated exactly and the intersite interactions are decoupled within the MFA (site-dependent):
\begin{equation}
\hat{n}_{i}\hat{n}_{j} \rightarrow \left\langle \hat{n}_{i} \right\rangle \hat{n}_{j} + \left\langle \hat{n}_{j}\right\rangle \hat{n}_{i} - \left\langle \hat{n}_{i}\right\rangle \left\langle \hat{n}_{j}\right\rangle.
\end{equation}
A~variational Hamiltonian for the model~(\ref{row:1}) has a form
\begin{equation}
\hat{H}_{0}=\sum_i{\hat{H}_i}=\sum_i{\left[U\hat{n}_{i\uparrow}\hat{n}_{i\downarrow}-\mu_i \hat{n}_i -\frac{1}{2} n_i\psi_i\right]},
\end{equation}
where \mbox{$\psi_i = \sum_{j\neq i}{W_{ij}n_{j}}$}, \mbox{$\mu_{i} =  \mu -\psi_i$} and \mbox{$n_i=\langle\hat{n}_i\rangle$}. $\hat{H}_{0}$
is diagonal in representation of occupancy numbers (i.~e. $\hat{H}_i$ is diagonal in the base consisting of \mbox{$|0\rangle$}, \mbox{$|\uparrow\rangle$}, \mbox{$|\downarrow\rangle$}, \mbox{$|\uparrow\downarrow\rangle$} at $i$ site with eigenvalues: $0$, doubly degenerated $-\mu_i$, and \mbox{$U-2\mu_i$}, respectively) and a~general expression for the free energy $F$ in the grand canonical ensemble in the VA is
\begin{equation*}
F = -\frac{1}{\beta}\ln\left\{\textrm{Tr}\left[\exp(-\beta\hat{H_{0}})\right]\right\} + \mu \langle \hat{N}_{e} \rangle,
\end{equation*}
where \mbox{$\beta=\frac{1}{k_{B}T}$}, \mbox{$\hat{N_{e}}=\sum_{i}{\hat{n}_{i}}$},
\mbox{$\langle\hat{N_{e}}\rangle=nN$} is the number of electrons in the system. The average value of operator $\hat{A}$ is defined as
\mbox{$\langle\hat{A}\rangle =
\frac{\textrm{Tr}\left[\exp(-\beta\hat{H}_{0})\hat{A}\right]}{\textrm{Tr}\left[\exp(-\beta\hat{H}_{0})\right]}$}.
\mbox{$\textrm{Tr}\hat{B}$} means a~trace of any operator $\hat{B}$ and it is calculated in the Fock space.

The explicit formula for the free energy obtained in the VA has the following form
\begin{equation}\label{row:sitedependentenergy}
F = \sum_{i}{\left\{n_i(\mu - \frac{1}{2}\psi_i) - \frac{1}{\beta}\ln{Z_i}\right\}},
\end{equation}
where
\begin{equation*}
Z_i = 1+2\exp{[\beta(\mu-\psi_i)]} + \exp{[\beta(2\mu - 2\psi_i - U)]},
\end{equation*}
while the expression for the average number of electrons at $i$-site is given by
\begin{equation}\label{row:nonsite}
n_i=\frac{2}{Z_i}\{\exp{[\beta(\mu - \psi_i )]}+ \exp{[\beta(2\mu - 2 \psi_i -U)]}\},\quad
\end{equation}
so one has a~set of \mbox{$N+1$} self-consistent equations to solve consisting of $N$ equations in form of~(\ref{row:nonsite}) (for every site from $N$ sites) and the condition (\ref{row:2}) in the form:
\begin{equation}\label{row:condition}
n-\sum_i{n_i}=0.
\end{equation}

The double occupancy $D_i$ of the site $i$ is determined by the following equation:
\begin{equation}\label{row:Dsitedependent}
D_i=\langle \hat{n}_{i\uparrow}\hat{n}_{i\downarrow}\rangle
=\frac{1}{Z_i} \exp{[\beta(2\mu - 2 \psi_i -U)]}.
\end{equation}

The solutions of the set \mbox{(\ref{row:nonsite})--(\ref{row:condition})} can correspond to a~minimum, a~maximum or a~point of inflection of the free energy (\ref{row:sitedependentenergy}) on the \mbox{($N-1$)-dimensional} manifold in \mbox{$N$-dimensional} space \mbox{$\{n_i\}_{i=1}^{N}$} defined by the condition (\ref{row:condition}). To find the solutions corresponding to stable (or metastable) states of the system, one should find a~conditional minimum of \mbox{$F$} with respect to all $n_i$ with the condition (\ref{row:condition}).

One can prove that for two-sublattice orderings, if \mbox{$F=F(n_A,n_B)$} have a~conditional minimum with respect to $n_A$ and $n_B$ with condition (\ref{row:condition}), then \mbox{$F=F(n,n_Q)$} have also a~minimum with respect to \mbox{$n_Q=(n_A-n_B)/2$} (if $n$ is fixed). So the procedure used in section~\ref{sec:ham} does not lose stable (metastable) solutions. In this instance, $F$ is the free energy only of homogeneous phases (as an~assumption) and cannot describe any phase separated states, which energies are calculated from (\ref{row:freeenergyPS}). One needs to check the stability condition \mbox{$\partial\mu/\partial n > 0$} for the homogeneous phases, which is one of the sufficient conditions for the conditional minimum of $F$.
In the case of two-sublattice orderings on the alternate lattices the \mbox{Eqs.~(\ref{row:sitedependentenergy})--(\ref{row:nonsite})} reduce to \mbox{Eqs.~(\ref{row:freeenergy})--(\ref{row:set2})} obtained in section~\ref{sec:ham}.


\end{document}